\documentclass{aastex63}

\shorttitle{A First Ascent M III -- NS Binary}
\shortauthors{Hinkle et al.}

\begin{document}

\title{ON ITS WAY TO THE NEUTRON STAR -- WHITE DWARF BINARY GRAVEYARD,\\  
IGR~J16194$-$2810, A FIRST ASCENT M GIANT X-RAY BINARY}

\author[0000-0002-2726-4247]{KENNETH H. HINKLE}
\affil{NSF's National Optical-Infrared Astronomy Research Laboratory,\\
P.O. Box 26732, Tucson, AZ 85726, USA}
\email{ken.hinkle@noirlab.edu}

\author[0000-0002-9413-3896]{FRANCIS C. FEKEL} 
\affil{Center of Excellence in Information Systems, Tennessee State University, \\
3500 John A. Merritt Boulevard, Box 9501, Nashville, TN 37209, USA}

\author[0000-0002-5514-6125]{OSCAR STRANIERO}
\affil{INAF, Osservatorio Astronomico d'Abruzzo,\\
I-64100 Teramo, Italy, and INFN, Sezione di Roma La Sapienza,  Roma, Italy}

\author[0000-0002-0475-3662]{ZACHARY G. MAAS}
\affil{Indiana University Bloomington, Astronomy Department,\\ 
727 East Third Street, Bloomington, IN 47405, USA}

\author[0000-0003-0201-5241]{RICHARD R. JOYCE}
\affil{NSF's National Optical-Infrared Astronomy Research Laboratory,\\
P.O. Box 26732, Tucson, AZ 85726, USA}

\author[0000-0002-0702-7551]{THOMAS LEBZELTER}
\affil{Department of Astrophysics, University of Vienna,
T{\"u}rkenschanzstrasse 17, 1180 Vienna, Austria}

and

\author[0000-0001-8455-4622]{MATTHEW W. MUTERSPAUGH}
\affil{Columbia State Community College, 1665 Hampshire Pike, Columbia, TN
39401, USA}

\author[0000-0003-3480-0957]{JAMES R. SOWELL}
\affil{School of Physics, Georgia Institute of Technology, Atlanta, GA 30332, USA}

\begin{abstract}
A single-lined spectroscopic orbit for the M giant in the X-ray binary 
IGR~J16194$-$2810 is determined from a time-series of optical 
spectra.  The spectroscopic orbital period of 192.5 days is twice 
that of the photometric period, confirming that the M giant in the 
system is an ellipsoidal variable. The giant is identified as a first 
ascent giant approaching the red giant tip.  The primary is a neutron 
star (NS) with its M giant companion filling its Roche lobe verifying 
the system classification as a Low-Mass X-ray binary (LMXB). Stellar 
C, N, O and Fe abundances are derived for the M giant with the 
C, N, O values typical for a field giant with [Fe/H] = $-$0.14. 
The system does not have a large kick velocity.  Models for the 
evolution of the system into a binary NS -- white dwarf (WD) are 
presented. The X-ray properties are discussed in the context of this 
model. This binary is a rare example of a luminous, long orbital period,  
LMXB early in the transient ellipsoidal phase.    
\end{abstract}


\section{INTRODUCTION}\label{introduction}

Late-type giants in binary systems with orbits of a few years or less 
are classified into one or more families of objects. Among the first 
categories to be identified was the `symbiotic' (SySt) class 
where an emission line spectrum is combined with a late-type 
giant spectrum \citep{merrill_humason_1932}.  While SySt are now 
known to be systems with mass loss from a giant being accreted by a 
degenerate secondary, neither the binarity nor the defining mass 
transfer was understood until decades after the discovery. SySt are 
now known to be part of a larger overall class of giants with a 
degenerate companion. Many of the other groups have peculiar 
abundances evidencing previous mass transfer. The orbital periods 
typically require evolution through a main sequence (MS) -- giant 
common envelope (CE) stage \citep{escorza_et_al_2020}.
While many of these systems are detached, there are SySt with
Roche lobe overflow (RLO).  Low mass RLO systems have ellipsoidal 
light variations and are members of the sequence E variables, 
a group of stars identified in surveys due to their period-luminosity 
relation \citep{wood_et_al_1999}. 

With the development of X-ray techniques, additional groups of 
related objects have been identified including the low mass X-ray 
binaries (LMXBs) and the X-ray symbiotics (SyXBs), both with a NS 
primary. In addition, SySt with WD primaries are now being found 
using X-ray surveys with many of these systems lacking the hallmark 
strong optical emission \citep{munari_2019}.  This is similar to the 
SyXBs where most are optically quiescent systems, having few, if any, 
optical emission lines.  Membership in the SySt group is based on 
morphology and not evolutionary commonality. For instance, for the 
SyXB, in addition to a previous CE stage, a supernova (SN) is 
required to create the NS \citep{verbunt_van_den_heuvel_1995}.

Nearly all of the SyXBs were first cataloged as X-ray sources that 
were then identified with late-type giants, thus LMXBs, and only 
subsequently classified as SyXBs.  The LMXB group consists 
mainly of slowly evolving NS -- MS binaries 
with systems containing a late-type giant making up a small fraction.
In nearly all LMXBs the mass donor 
star is in RLO \citep{verbunt_1993}.  The RLO systems evolve into 
NS -- WD binaries and, in the process, the transfer of angular 
momentum spins up the NS. Millisecond pulsar NS -- WD binaries are 
astrophysically important, and there is extensive literature 
(see, for instance, reviews by \citet{bhattacharya_van_den_heuvel_1991},
\citet{phinney_kulkarni_1994}, and \citet{verbunt_van_den_heuvel_1995}). 

In this paper we discuss the binary IGR~J16194$-$2810. This star is
an excellent example of inclusion in multiple families of binaries.  
Initially identified as a NS or black hole (BH) X-ray source 
\citep{bird_et_al_2006,bassani_et_al_2006,stephen_et_al_2006}, 
IGR~J16194$-$2810 was later found to be coincident with an M III star
\citep{masetti_et_al_2007}, and hence, was a LMXB. From time-series 
survey photometry, \citet{kiraga_2012} found the M III star to be an 
ellipsoidal variable. Due to the M III companion to a NS, it has also 
been included in lists of SyXB \citep{yungelson_et_al_2019}.
The multiple classifications do not give insight into the significance 
of this system.  It stands out in the various lists of objects because 
it is ellipsoidal, has a period that is both abnormally short for a 
SyXB/SySt and unusually long for a LMXB, but does not have a high 
X-ray luminosity. We have undertaken a program of spectroscopy to 
explore the nature of this highly unusual binary.

In the following sections we discuss the optical and near-infrared 
spectroscopy of the binary system IGR~J16194$-$2810. The new 
observations are reviewed in the next section. In Section 
\ref{sec:orbit} we discuss the orbit. Section 
\ref{sec:parameters_section} starts with a discussion of the 
identification of IGR~J16194$-$2810 as a NS binary 
followed by a detailed discussion of the properties of the late-type 
companion, including the space motion, effective temperature, 
luminosity, and radius. The mass range of the NS is also reviewed. 
Then in Section \ref{sec:ellip_rot_radius} we use its ellipsoidal 
properties to extract limits on the NS and M III masses and binary 
inclination. In Section \ref{sec:abund}, the M giant abundance 
determinations are discussed. Section \ref{sec:discussion} 
includes a discussion of the binary evolution of IGR~J16194$-$2810 
with a focus on the end product. Comparisons with various related 
groups of variables are included, as is a new model based on the 
current parameters of the system.  A summary is presented in the 
Conclusions, Section \ref{sec:conclusion}.


\section{OBSERVATIONAL DATA}\label{sec:obs}

\subsection{Near-IR Spectra}


The $H$ and $K$ regions were observed at spectral resolution $R
~=~\lambda/\Delta\lambda$ of $\sim$45000 on 2018 April 22 UT 
with the Immersion Grating Infrared Spectrometer
\citep[IGRINS,][]{park_et_al_2014} at the Gemini South 8-meter 
telescope (GS).  Reductions were the same as those employed for 
other SyXB spectra observed with IGRINS/GS \citep{hinkle_et_al_2019}. 
In brief, the initial reduction was done with the IGRINS pipeline.  
The output from this process is the echelle orders ratioed
to a telluric reference standard.  The continuum in each order was
then normalized by doing a linear fit to the high points.  
Polynomial terms in the continuum were removed with the 
IRAF\footnote{Image Reduction and Analysis Facility software 
distributed by NOAO (now NOIRLab).} continuum routine $splot~'t'$ 
at low order. The orders were joined by matching the overlap 
regions between the orders. The velocities of the $K$-band CO 
first overtone lines were measured, and the average velocity
is listed in Table \ref{table:rv_obs}.

Nearly a year after the IGRINS observation, an observation of the 
2.299 -- 2.311 $\mu$m region of the $K$ band was taken on 
2019 March 17 UT with the Phoenix cryogenic echelle spectrograph 
\citep{hinkle_et_al_1998} on GS.  The spectrum has $R~\approx~50000$.
The spectral orders were extracted from the raw data and wavelength 
calibrated with standard reduction techniques \citep{joyce_1992}.  
In late-type stars the spectral region observed is dominated by a 
series of strong CO 2-0 lines.  The stellar velocity was measured 
by cross correlation of CO laboratory frequencies to the spectrum 
with the IRAF cross-correlation routine $fxcor$ \citep{fitzpatrick_1993}.

The Phoenix spectrum has much less wavelength coverage than the 
IGRINS spectrum but proved to be a critical observation. Comparison 
of the IGRINS and Phoenix radial velocities showed a change
consistent with orbital motion at the ellipsoidal period of 
192.8 days \citep{kiraga_2012}. No changes in the spectrum resulting 
from late-type stellar variability were apparent.  On the basis of 
this observation, an effort was launched to obtain a spectroscopic 
orbit of the M III.

\subsection{Visual Spectra}

A time series of optical high-resolution spectra was obtained at 
Fairborn Observatory in southeast Arizona with the Tennessee State 
University 2~m Automatic Spectroscopic Telescope (AST) and 
fiber-fed echelle spectrograph \citep{eaton_williamson_2004, 
eaton_williamson_2007}. The detector is a Fairchild 486 CCD that has 
a 4096 $\times$ 4096 array of 15 $\mu$m pixels \citep{fekel_et_al_2013}. 
There are 48 echelle orders that range in wavelength from 3800 to 8260~\AA. 
The observations were obtained with a 365 $\mu$m fiber that produces 
a resolution of 0.4~\AA\ or $R~\approx~15000$ at 6000~\AA. 

With a $V$ magnitude of 12.75 \citep{kiraga_2012} and a declination of 
$-$28\arcdeg, IGR~J16194$-$2810 is near the limit of observability of 
the AST. In addition, due to its southern declination, the observing 
season for IGR~J16194$-$2810 is relatively short. In spite of these 
difficulties, between 2019 June and 2022 June, 63 useful spectroscopic 
observations were acquired. 


A discussion of velocity measurement for AST spectra can be found 
in \citet{fekel_et_al_2009}. While IGR~J16194$-$2810 is grouped with 
the symbiotic stars, as with a number of other SyXB, the optical 
spectrum is that of a normal late-type star, i.e., it does not 
contain conspicuous emission lines or extensive veiling caused by 
continuum emission. Thus, the reference-star line list that was 
used for measurement contains 40 lines that are relatively unblended 
in M giant spectra and range in wavelength from 5000 to 6800~\AA.  
To fit individual line profiles, a rotational broadening
function was used \citep{lacy_fekel_2011, fekel_griffin_2011}.  
With the same reduction technique, unpublished velocities of several 
IAU radial-velocity standards observed with the 2~m AST have an 
average velocity difference of $-$0.6 km~s$^{-1}$ when compared to 
the results of \citet{scarfe_2010}. Thus, we have added 0.6 
km~s$^{-1}$ to each of our AST velocities. Our radial-velocity 
observations are listed in Table~\ref{table:rv_obs}.

In high-resolution spectra the IGR~J16194$-$2810 lines are 
conspicuously broadened. From a dozen spectra with the best 
signal-to-noise ratios, the rotational broadening fits resulted in 
a projected rotational velocity, $v\,sin\,i$, of $16 \pm 1$
km~s$^{-1}$ where $v$ is the rotational velocity and $i$ is 
the inclination. The technique used to measure $v\,sin\,i$ is 
described in \citet{willmarth_et_al_2016} and 
\citet{henry_et_al_2022}. Spectrum synthesis of the much 
higher signal-to-noise IR spectra require a projected rotational 
broadening of 17 km s$^{-1}$, a value in accord with the optical 
measurements. 

\subsection{Photometry}

\begin{figure}
\centering
\includegraphics[width=.8\linewidth]{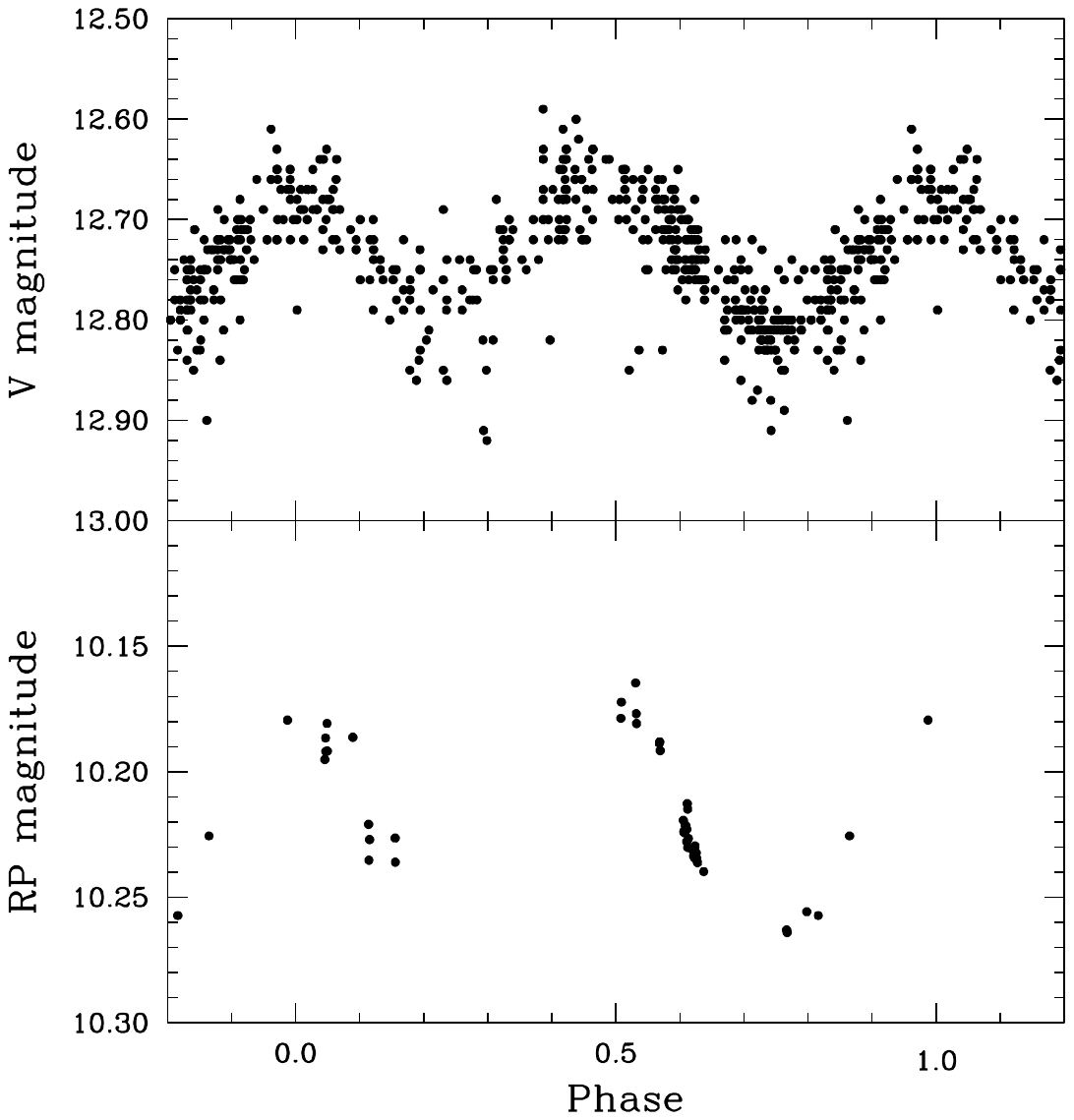}
\caption{Top: 
The $V$ band light curve of IGR~J16194$-$2810 phased to 192.8 d
from \citet{kiraga_2012}.
Bottom: The Gaia $RP$-band light curve of IGR~J16194$-$2810. 
The system is ellipsoidal and there are two maxima and minima 
in each orbital cycle.
}
\label{fig:light_curve}
\end{figure}

An $I$-band light curve for IGR~J16194$-$2810 is presented by 
\citet{kiraga_2012}. The photometry, 375 observations, is from the 
All Sky Automated Survey (ASAS).  At the magnitude of IGR~J16194$-$2810, 
the uncertainty of an individual measurement can be as much as several 
tenths of a magnitude but is compensated by the large number of 
observations. IGR~J16194$-$2810 is, in fact, fainter than the cutoff 
magnitude of the \citet{kiraga_2012} survey. A discussion of the ASAS 
data, which includes the IGR~J16194$-$2810 photometry, is given by 
\citet{pojmanski_maciejewski_2004}. Briefly, independent instruments 
equipped with standard $V$, $R$, and $I$ filters took simultaneous 
photometry through five different sized apertures. For IGR~J16194$-$2810 
only $V$ and $I$ data are available. The light curve amplitude of 
IGR~J16194$-$2810 is $\sim$0.11 mag at $I$ and $\sim$0.14 mag at $V$.
The $V$ data are shown in the upper panel of Figure \ref{fig:light_curve}.  
Higher precision photometry is available from Gaia but with much less 
frequent sampling.  Gaia $RP$-band \citep{jordi_et_al_2010} observations 
are shown in the lower panel of Figure \ref{fig:light_curve}.


\section{SPECTROSCOPIC ORBIT}\label{sec:orbit}


\citet{kiraga_2012} proposed an orbital period of 192.8 days 
for IGR~J16194$-$2810 based on the ASAS photometric observations. 
From the shape and amplitude of the light curve, \citet{kiraga_2012}
assigned IGR~J16194$-$2810 to the ellipsoidal class of variables.
For ellipsoidal variables the photometric period is half the orbital 
period, so for IGR~J16194$-$2810 the photometric period is 96.4 days 
(Figure \ref{fig:light_curve}).  Ellipsoidal variations result from 
the axial elongated shape of the star(s) with the light curve 
repeating as the long and short sides of the star are seen twice 
during an orbit \citep{morris_1985}.

To compute the orbit, equal weights were assigned to the lone IGRINS 
and Phoenix radial velocities as well as each AST 
velocity. Adopting the 192.8 day period as a starting value, we 
obtained a preliminary orbital solution with the computer 
program BISP \citep{wolfe_et_al_1967}, which uses the Wilsing-Russell 
method \citep{wilsing_1893, russell_1902}. We next refined those 
elements with the program SB1 \citep{barker_et_al_1967}, which 
utilizes differential corrections. That solution produced a period 
of 192.6 $\pm$ 0.2 days and an eccentricity of 0.018 $\pm$ 0.008. 
Given the very small eccentricity, we next obtained a circular orbit 
solution with the SB1C orbit program (D. Barlow 1998, private 
communication), which iterates sine/cosine fits by differential 
corrections. That solution resulted in a period of 192.5 $\pm$ 0.2 
days.  The orbital and photometric periods are the same to within the 
uncertainty. 

In the case of IGR~J16194$-$2810, the precepts of \citet{lucy_sweeney_1971} 
for discriminating between an eccentric and a circular orbit indicate 
that the eccentric orbit is to be preferred. However, the phase 
distribution of velocities is far from uniform with much of the 
drop from maximum velocity to minimum velocity missing. Despite 
this lack of velocities in that part of the orbit, the resulting 
orbital eccentricity is only 0.018 with a longitude of periastron,
$\omega$, of 64\fdg4 $\pm$ 20\fdg1.  \citet{lucy_sweeney_1971} have 
argued that such an eccentricity value is so small that it is likely 
statistical in origin and therefore, not significant. Thus, we have 
chosen to adopt the circular orbit for IGR~J16194$-$2810. 

While the eccentricity that we compute is statistically insignificant,
improvements in the orbit may ultimately find a very slight 
eccentricity.  \citet{sterne_1941} noted that small eccentricities 
can result from the elliptical distortion of the stars in 
ellipsoidal systems. For the ellipsoidal SySt T~CrB,
\citet{kenyon_garcia_1986} argued for the reality of small 
eccentricities.  This work has been reviewed and corrected by 
\citet{belczynski_mikolajewska_1998}. For an orbital eccentricity 
resulting from ellipsoidal distortions the longitude of periastron, 
$\omega$, is expected to be 90$^\circ$ or 270$^\circ$. For T~CrB, 
\citet{kenyon_garcia_1986} find $\omega = 80^\circ\pm6^\circ$.  
In our IGR~J16194$-$2810 orbit, $\omega$ = 64$^\circ$ with a large 
uncertainty of 20$^\circ$.

\begin{figure}
\centering
\includegraphics[width=.8\linewidth]{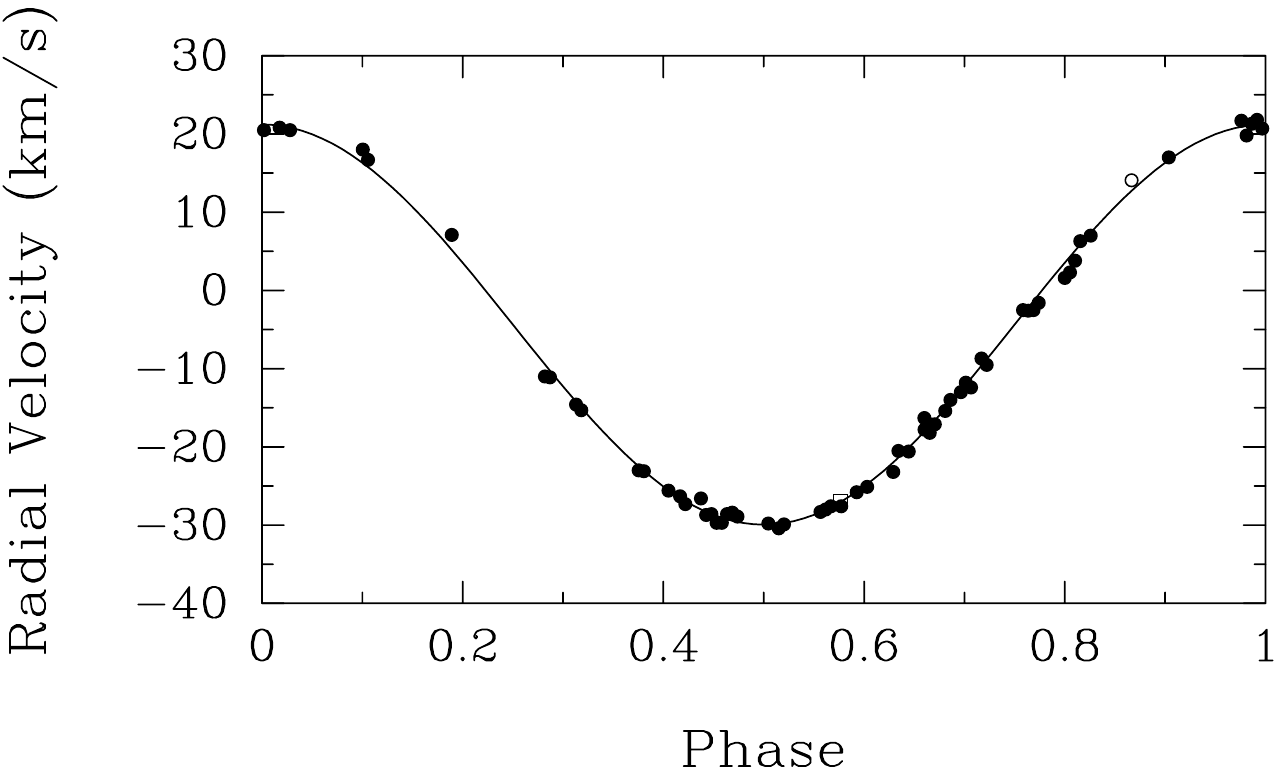}
\caption{
A plot of the IGR~J16194$-$2810 radial velocities (Solid circles = 
Fairborn Observatory, open circle = IGRINS, and open box = Phoenix) 
compared with the computed velocity curve (solid line). 
The standard error of an observation is 0.9 km s$^{-1}$. 
Phase zero is a time of maximum velocity.  The binary has an orbital period of
192.5 days.
}
\label{fig:orbit_rvs}
\end{figure}

The orbital elements and derived parameters for IGR~J16194$-$2810 are 
listed in Table~\ref{table:orbit}. In a circular orbit the element 
$T$, a time of periastron passage, is indeterminate. So, as 
recommended by \citet{batten_et_al_1989}, $T_0$, a time of maximum radial 
velocity is given instead.  A phase plot of the radial velocities 
compared with the computed orbital curve is shown in 
Figure~\ref{fig:orbit_rvs}, where phase zero is a time of maximum 
velocity.


\section{STELLAR PARAMETERS}\label{sec:parameters_section}


Since IGR~J16194$-$2810 is an ellipsoidal variable, the stellar 
parameters can be derived in a variety of ways. In the following 
subsections we discuss the various inputs and resulting values. 
The parameters derived are summarized in Table \ref{table:parameters}.


\subsection{Identification and Spectral Type}\label{sec:history}

The X-ray source IGR~J16194$-$2810 was first detected in X-rays above 
20 keV by $INTEGRAL/IBIS$\footnote{INTErnational Gamma-Ray Astrophysics 
Laboratory/Imager on Board the INTEGRAL Satellite} 
\citep{bassani_et_al_2006,bird_et_al_2006}. \citet{stephen_et_al_2006} 
identified that source with $ROSAT$\footnote{ROentgen SATellite} 
1RXS~J161933.6$-$280726 and with a $B$$\sim$14 mag optical counterpart. 
No pulsations have been detected in the X-ray flux 
\citep{enoto_et_al_2014}. Additional directed observations by 
\citet{masetti_et_al_2007} from 
$Swift/XRT$\footnote{$Swift$/X-Ray Telescope} improved the coordinates 
clearly identifying the M2~III star, 2MASS~16193334$-$2807397, as the 
optical source.  As a result, IGR~J16194$-$2810 is a member of the 
LMXB class.

\citet{ratti_et_al_2010} confirmed the identification using $Chandra$.
A NS in the IGR~J16194$-$2810 binary, originally proposed by 
\citet{ratti_et_al_2010}, was supported by \citet{kitamura_et_al_2014}.  
They found that the hard X-ray emission originates in a small,
0.7 km, radius on the NS surface with the soft emission possibly from 
the accretion stream. In addition, the X-ray flux and hardness confirm 
the NS nature of the degenerate in the IGR~J16194$-$2810 system.
The X-ray luminosity is more than 100 times the luminosity of typical 
WD SySt \citep{luna_et_al_2013}.

The 3850--7200 \AA\ region of the spectrum, observed at R$\sim$1000,  
is shown in Figure 2 of \citet{masetti_et_al_2007}. The spectral 
type assigned, M2~III, is based largely on the near-IR TiO bands.
The conspicuous presence of TiO in the spectrum requires that the 
giant be of type M \citep{morgan_et_al_1943,abt_et_al_1968}.
\citet{pecaut_mamajek_2016} independently found a spectral type M2~III 
based on a different data set.   The optical spectrum of IGR~J16194$-$2810 
does not contain emission lines, which are a hallmark of traditional 
SySt systems.  

\begin{figure}
\centering
\includegraphics[width=.8\linewidth]{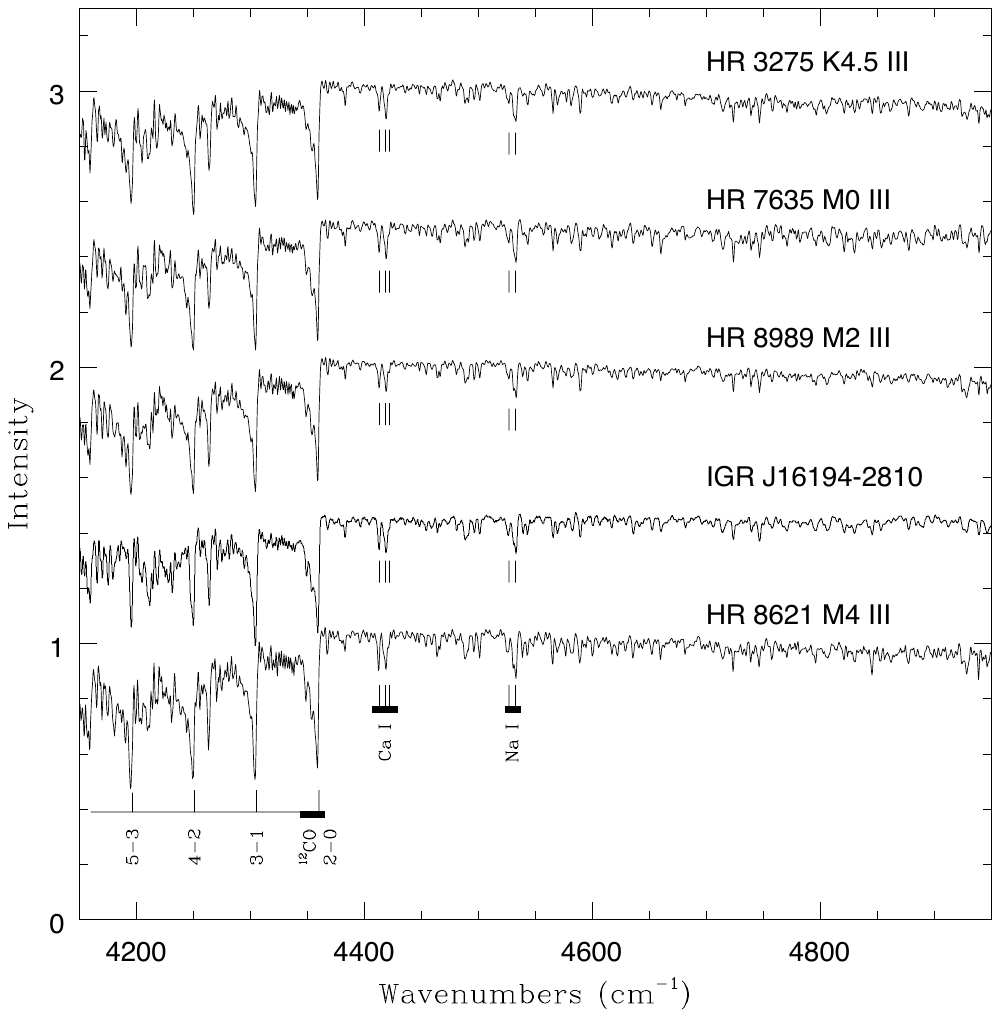}
\caption{
The IGRINS $K$ band spectrum of IGR~J16194$-$2810 convolved to R=3000 and
compared to standard giant star spectra from \citet{wallace_hinkle_1997}.
The bars
mark the \citet{ramirez_et_al_1997} spectral indices. 
The major spectral features are indicated
by vertical lines.
}
\label{fig:K_band_spectra}
\end{figure}
\begin{figure}
\centering
\includegraphics[width=.8\linewidth]{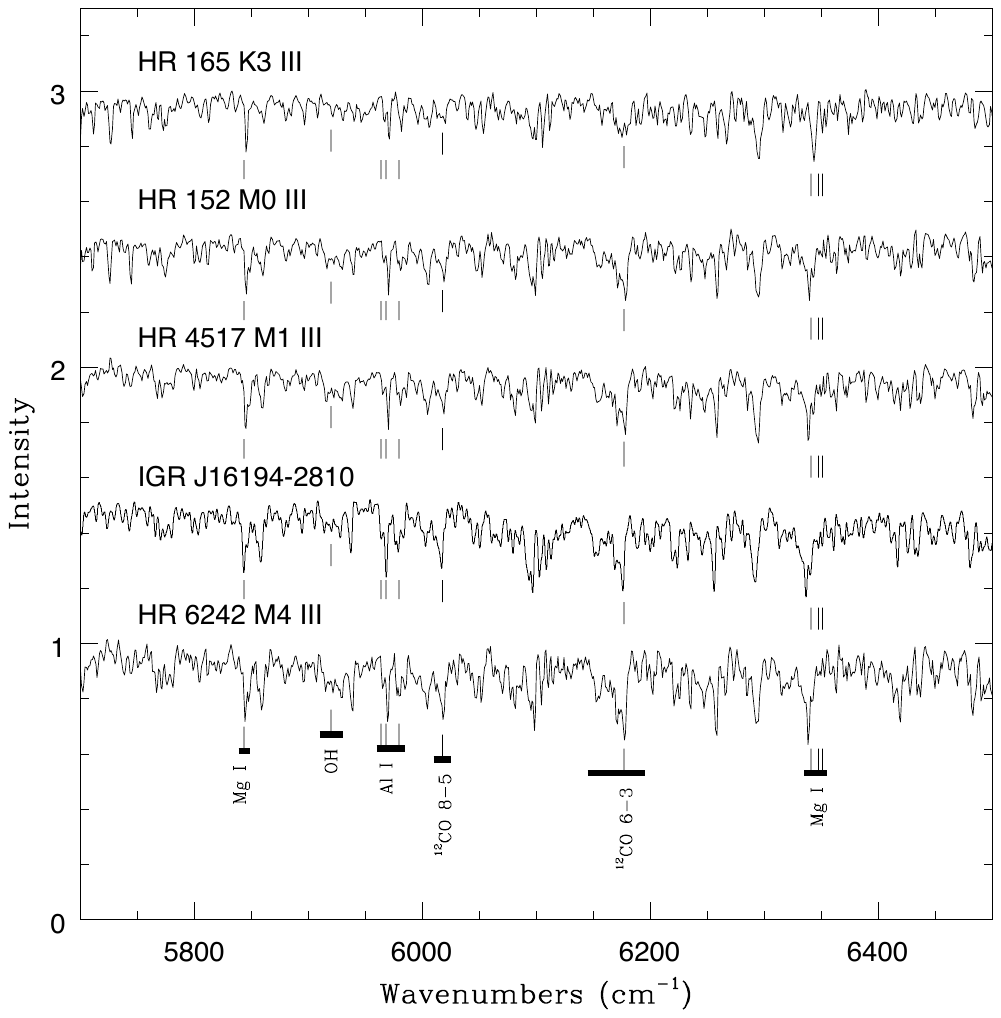}
\caption{
The IGRINS $H$ band spectrum of IGR~J16194$-$2810 convolved to R=3000 and
compared to standard giant star spectra from \citet{meyer_et_al_1998}. The bars
mark the cool star spectral indices of \citet{meyer_et_al_1998} with 
the spectral features listed in Table 4 of \citet{meyer_et_al_1998} indicated
by vertical lines.  The increase in strength of the $^{12}$CO 6-3 index 
with cooler spectral type is apparent in the Figure. 
}
\label{fig:H_band_spectra}
\end{figure}

The IGRINS spectrum covers the $K$ and $H$ bands.  The 
spectrum was convolved to R=3000 for use in spectral typing.  
The $K-$ and $H$-spectra are shown in Figures \ref{fig:K_band_spectra} 
and \ref{fig:H_band_spectra} along with 
M giant standard star
spectra \citep{wallace_hinkle_1997,meyer_et_al_1998}.  Visual comparison confirms 
that the near-IR spectrum of IGR~J16194$-$2810 is that of a normal, 
early M~III star but provides no refinement of the spectral type.  


\subsection{Distance and Space Motion}\label{sec:space_vel}

The Gaia data release 3 (DR3) parallax for IGR~J16194$-$2810 
is 0.452 $\pm$ 0.036 mas, corresponding to a distance of 
2101$_{-150}^{+129}$ pc \citep{bailer-jones_et_al_2021}.
Previous distance estimates from extinction and X-ray properties 
were in the range 3.0 -- 3.7 kpc 
\citep{ratti_et_al_2010,masetti_et_al_2007}. The X-ray luminosity 
($L_X$) corrected to 2.1 kpc, $2 \times 10^{35}$ ergs s$^{-1}$, 
is similar to that of other SyXB \citep{masetti_et_al_2007}.  

\citet{ratti_et_al_2010} found a proper motion of $1.3\pm4.7$ 
mas~yr$^{-1}$ in right ascension and $-20.2\pm4.7$ mas~yr$^{-1}$ in 
declination and concluded that IGR~J16194$-$2810 has a minimum peculiar 
velocity of $280 \pm 66$ km s$^{-1}$, suggesting a kick velocity.
The Gaia proper motion, $-0.894 \pm 0.043$ mas~yr$^{-1}$ in right
ascension and $-5.287 \pm 0.027$ mas~yr$^{-1}$ in declination 
\citep{gaia_2022}, is quite different, as well as two orders of 
magnitude more accurate.

\begin{figure}
\centering
\includegraphics[width=0.8\linewidth]{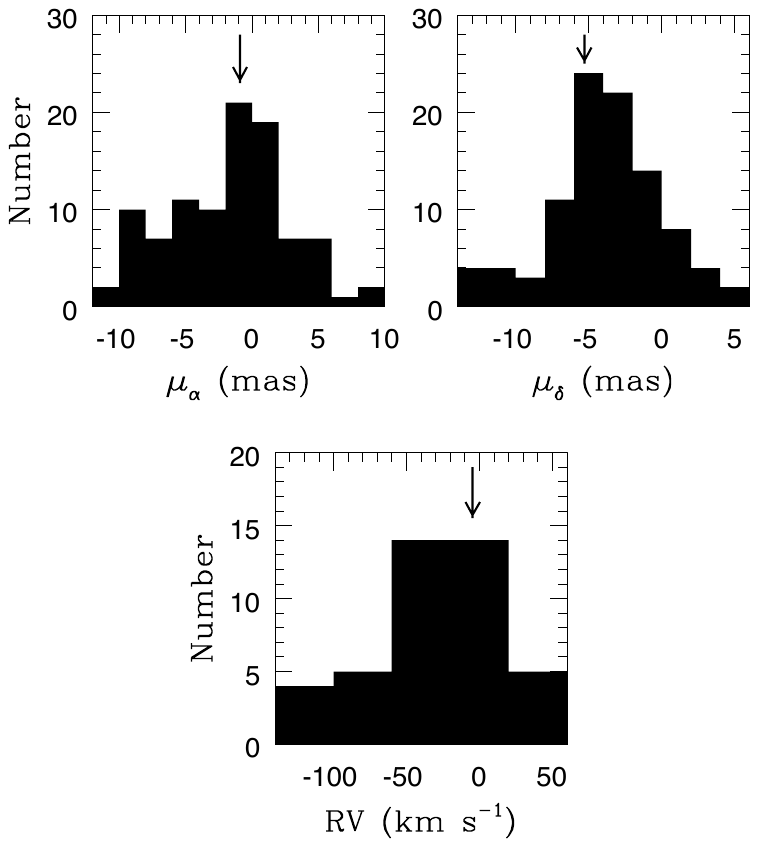}
\caption{Histograms of Gaia DR3 proper motions and radial velocities 
in 20 arcmin radius around the position of IGR~J16194$-$2810.  The 
parallax was restricted to between 0.35 and 0.65 mas with a parallax 
error less than or equal 0.045 mas. The positions of the 
IGR~J16194$-$2810 values are marked with the arrows.
}
\label{fig:space_motion}
\end{figure}

Using a beam size of circular radius 20 arcmin centered on the position 
of IGR~J16194$-$2810, sources in the Gaia DR3 archive with parallaxes 
between 0.35 and 0.65 mas and having a parallax error $\leq$0.045 mas 
were selected. This results in a total of 98 stars.  Histograms of the 
proper motion of these sources are shown in Figure \ref{fig:space_motion}.  
In this selection of DR3 data, 42 sources also have measured radial 
velocities, and a histogram of those velocities is also in Figure 
\ref{fig:space_motion}. The arrows mark the values for IGR~J16194$-$2810.  
The Gaia data shows that IGR~J16194$-$2810 does not have a significant 
kick velocity.  With the assumption that the parallax and proper 
motion uncertainties are independent, the transverse velocity is 
$-$8.9 $\pm$ 1.0 km s$^{-1}$ in right ascension and 
$-$52.7$_{-3.5}^{+4.0}$ km s$^{-1}$ in declination.
The systemic ($\gamma$) radial velocity of the binary is
$-$4.36 $\pm$ 0.13 km s$^{-1}$ (Table \ref{table:orbit}).
These results produce a space velocity of $-$57.7 km s$^{-1}$.
This value appears typical for stars in this region of
the Milky Way (Figure \ref{fig:space_motion}).


\subsection{Effective Temperature, Luminosity, Radius}\label{sec:temp_lum}

Multiple techniques are available for determining the effective 
temperature.  We employ two, one based on optical/near-IR 
spectroscopy and the other on infrared photometry. As mentioned in 
Section \ref{sec:history}, optical spectra of IGR J16194$-$2810 
are classified as spectral type M2~III based on TiO bands in the red.  
The effective temperature for a M2~III is $\sim$3700 K 
\citep{dyck_et_al_1996,van_belle_et_al_1999}. It is difficult to set 
an uncertainty on this value due to uncertainty in the spectral 
classification itself, uncertainty in the spectral types of 
the reference spectra, and the known scatter in the calibration of
effective temperature against other parameters 
\citep[e.g., see][]{ramirez_et_al_1997}.


A more quantified approach is possible with the use of spectral 
indices. Several authors have calibrated spectral indices in 
medium resolution $K$- and $H$-band spectra. With the use of 
empirical relations, the indices can be converted to effective 
temperatures with uncertainties. \citet{ramirez_et_al_1997} 
discuss three $K$-band indices, a Na blend at 4524 -- 4535 cm$^{-1}$, 
a Ca blend at 4409 -- 4426 cm$^{-1}$, and the $^{12}$CO 2-0 
bandhead from 4344 -- 4362 cm$^{-1}$. We measured the equivalent 
widths of the indices in \citet{wallace_hinkle_1997} K4 III to 
M4 III standard spectra and the IGRINS IGR J16194$-$2810 $K$-band 
spectrum convolved to R=3000 (Figure \ref{fig:K_band_spectra}).
The continuum points of the spectra 
were normalized. Our measurements of the Ca and Na indices in the 
M~III standard spectra have 
considerable scatter and are not useful 
for early M stars. \citet{ramirez_et_al_1997} found a similar result.  
However, also as found by \citet{ramirez_et_al_1997}, the CO index 
proved to be robust. Calibrating to the standard star equivalent 
widths listed in \citet{ramirez_et_al_1997}, the 2-0 $^{12}$CO 
bandhead index for IGR J16194$-$2810 has an equivalent width of 
3.79$\pm$0.10 cm$^{-1}$, 20.0$\pm$0.5 \AA, with its uncertainty 
mainly from the continuum placement. Combining uncertainties from 
the $T_{eff}$ calibration and measurement, $T_{eff}$ = 3660 $\pm$ 
190~K.


To utilize the spectral indices in the $H$-band, the high-resolution 
IGRINS spectrum was convolved to $R = 3000$ and normalized to the 
continuum of the reference standard star spectra.  The spectral indices
of \citet{meyer_et_al_1998} are marked on Figure
\ref{fig:H_band_spectra}.  
Three indices were recommended by \citet{meyer_et_al_1998} for measuring
the effective temperature of cool stars.
The OH 5920 cm$^{-1}$ index has an equivalent width of 1.01 $\pm$ 0.10 cm$^{-1}$ and 
that of 
the Mg~I 6345 cm$^{-1}$ index is 1.96 $\pm$ 0.10 cm$^{-1}$. For early 
M giants the values of these indices are relatively small and 
sensitive to the continuum value, and, as a result, the uncertainty 
is large. From expression 6 of \citet{meyer_et_al_1998} $T_{eff}~=~3690 
\pm$ 430 K. As in the $K$-band, the $H$-band CO index is more robust. 
The CO 6170 cm$^{-1}$ index has an equivalent width of 3.65 $\pm$ 
0.26 cm$^{-1}$, corresponding to $T_{eff}$ = 3660~K. An index -- effective 
temperature relationship with uncertainties is not provided by 
\citet{meyer_et_al_1998}. Based on the measurement uncertainty and the 
scatter in values for the standards, the total uncertainty is 
$\gtrsim\pm$200~K.

We next discuss two methods for determining the effective temperature 
using photometry. A catalog of effective temperatures for symbiotic 
stars based on fits to the 1 to 15 $\mu$m spectral energy 
distributions (SEDs) was produced by \citet{akras_et_al_2019}. The 
$T_{eff}$ for IGR J16194$-$2810 is 3779 $\pm$ 152~K.  However, the 
effective temperature depends on the reddening and 
\citet{akras_et_al_2019} did not take reddening into account.  We 
independently applied blackbody fitting to the same archival 
photometry and found an ambiguity between a $\sim$3700~K fit with 
no reddening and a marginally better fits at higher temperatures 
with enhanced reddening, in particular 4000~K with $A_V$ = 2.6 mag.
The catalog associated with \citet{akras_et_al_2019} also gives a 
Gaia effective temperature of 3511~K. The colors for IGR J16194$-$2810 
fall outside the color calibration for Gaia so this effective 
temperature is based on an extrapolation and is not valid.

To determine the reddening, the DUST database\footnote{ 
https://irsa.ipac.caltech.edu/applications/DUST/; 
\citet{schlafly_finkbeiner_2011}} was used to estimate the ISM 
contribution to the reddening. This indicates $A_V$ = 1.61 mag 
\citep{green_et_al_2018, green_et_al_2019}, corresponding to 
$A_K$ = 0.19 mag. Additional reddening due to circumbinary dust 
is possible but difficult to estimate. For IGR J16194$-$2810 the 
VizieR\footnote{https://vizier.cds.unistra.fr/viz-bin/VizieR; 
\citet{ochsenbein_et_al_2000}} catalog lists $K_s$ = 6.97 mag and 
Johnson $K$ and $J$ = 6.99 mag and 8.268 mag, respectively. With 
the use of standard reddening relations \citep{rieke_lebofsky_1985} 
and interstellar reddening with $A_V$ = 1.61 mag, $K_0$ = 6.79 mag
and $J_0-K_0$ = 1.01 mag. In the case of additional circumbinary reddening, 
assuming $A_V$=2.6 mag, $K_0$ = 6.60 mag and $J_0-K_0$ = 0.84 mag. 

Various color-color indices can be used to derive the effective 
temperature.  Assuming ISM reddening, the $J_0-K_0$ corresponds to 
$T_{eff} \sim$ 3700~K from the calibration of \citet{tokunaga_2000}. 
Similarly, the closest matching $J-K$ in the \citet{worthey_lee_2011} 
models of synthetic photometry are for a log($g$) = 0.5 and [Fe/H] = 0 
model and an effective temperature of 3735~K. For log($g$) = 1, 
$T_{eff}$ = 3672~K. Using the \citet{houdashelt_et_al_2000} models, 
the color matches $T_{eff}$ = 3700~K with log($g$) = 0.0 and [Fe/H] = 
+0.25. With their calibration this corresponds to spectral type M2.6~III.  
Additional reddening raises the effective temperature.  Adopting 
$J_0-K_0$ for circumbinary reddening, the effective temperature is 
$\sim$4400~K, typical of an early K giant. Since this is clearly not 
the case,  $A_V = $1.61 mag is adopted.  

The unweighted mean value from all the above approaches is 3700~K.  
While systematic uncertainties are present, we assume that the 
uncertainty is largely from measurement uncertainty.  In this case, 
the uncertainty on the mean is $\pm$100K. We adopt 3700 $\pm$ 100~K 
for the effective temperature of IGR J16194$-$2810.

The luminosity can be derived from the Gaia distance of 
2101$_{-150}^{+129}$ pc and photometry.  Using $(J_0-K_0) = 1.01$ mag,
\citet{bessell_wood_1984} give a bolometric correction of 2.71 mag.  
Combining this with the 
distance modulus of 11.61$^{+0.13}_{-0.16}$ mag and the
dereddened observed $K$ magnitude, $K_0$, of 6.79 mag,  
the absolute bolometric magnitude is -2.22$^{+0.17}_{-0.12}$.  The 
luminosity is 573$^{+72}_{-79}$ L$_\odot$.  Using the effective 
temperature -- radius -- luminosity relation and $T_{eff}$ = 3700 
$\pm$ 100~K, the radius is 58 $\pm$ 7 $R_\odot$. This is in 
agreement with the radius for a standard M2~III, 61 $\pm$ 11 $R_\odot$  
\citep[Table 7 of][]{van_belle_et_al_1999}. 


\subsection{Mass of the NS}\label{mass_ns}

Neutron stars have a limited range of mass from near the Chandrasekhar limit to 
near collapse into a black hole. A commonly made assumption is that 
a NS has a mass of 1.35 M$_\odot$ or 1.4 M$_\odot$ (see, for examples, 
references in the Australia Telescope National Facility (ATNF) Pulsar 
Catalog\footnote{\citet{manchester_et_al_2005};  
http://www.atnf.csiro.au/research/pulsar/psrcat}). A mass of 
$\sim$1.35 M$_\odot$ corresponds to both the Chandrasekhar limit and 
a sharp peak in the distribution of pulsar masses. While the mass
distribution of the overall population of NS is peaked at $\sim$1.36 
M$_\odot$, as discussed in \citep{ozel_freire_2016}, the pulsar 
population can be divided into groups.  Double NS pulsar binaries 
have masses of 1.33 $\pm$ 0.09 M$_\odot$.  Two peaks have been 
identified for millisecond pulsars at 1.49 $\pm$ 0.19 and 1.54 $\pm$ 
0.23 M$_\odot$. NS masses determined with reasonable accuracy range 
over $\sim$ 1.2 -- 2 M$_\odot$ \citep{ozel_freire_2016}.

It could be argued that the IGR~J16194$-$2810 NS falls into one of 
the groups identified by \citet{ozel_freire_2016}, for instance, 
that the evolutionary end stage of IGR~J16194$-$2810 is likely a 
NS--He WD. \citet{smedley_et_al_2014} find that to make the 
inclinations of known millisecond pulsar NS--He WD systems random, 
the mass of the NS must be in the range 1.55 -- 1.65 M$_\odot$.  
However, this not only assumes an end state for IGR~J16194$-$2810 but 
also that the system is at the end of the mass transfer 
phase. We start by assuming a mass for the NS that is in the full 
range of 1.2--2 M$_\odot$.  More restricted bounds on the NS mass 
are discussed below.


\section{ELLIPSOIDAL VARIATION}\label{sec:ellip_rot_radius}

A standard approach for determining the properties of ellipsoidal 
binaries employs a Wilson -- Devinney analysis of the light curve 
to determine the radius of the visible 
star, the semimajor axis, the inclination, and the mass ratio
\citep{wilson_devinney_1971}.  In 
the case of IGR~J16194$-$2810, these parameters can be solved with the
use of the mass function and $v\,sin\, i$ from the line profiles. The light curve 
provides additional information on limb darkening and gravity darkening coefficients 
\citep{morris_1985}.  Unfortunately, the uncertainty in the current 
photometry is too large to provide useful information.

\subsection{Mass Function and Rotation}\label{subsec:mass_fcn_rotation}

The star(s) in ellipsoidal variables are elongated due to Roche lobe 
filling. While photometry shows that the M2 III in IGR~J16194$-$2810 is 
clearly elongated, a critical question is how much of the Roche lobe 
is filled.  The light curve amplitude in the $I$ band, $0.11$ magnitudes, 
places IGR~J16194$-$2810 in 
the group with the largest amplitudes \citep{soszynski_et_al_2004}.  The discussion 
continues with the assumption that the red giant is Roche lobe filling.

For Roche lobe filling systems, the ratio of the masses 
of the stars, $q$, is a critical parameter.  We define 
$$ 
q~=~\rm{M}_\mathit{RG}/\rm{M}_\mathit{NS} 
$$
where the mass of the M2 III is M$_{RG}$ and the mass of the NS is M$_{NS}$. 

\citet{morris_1985} has shown that a complete solution for the 
parameters of an ellipsoidal system is possible for systems where the 
period, projected rotational velocity, light curve amplitude, spectral 
type, and mass of one of the components is known. Several approaches 
are possible. In this section we proceed with a graphical solution 
where the light curve amplitude is not required \citep{belczynski_mikolajewska_1998}.  

\begin{figure}
\centering
\includegraphics[width=.8\linewidth]{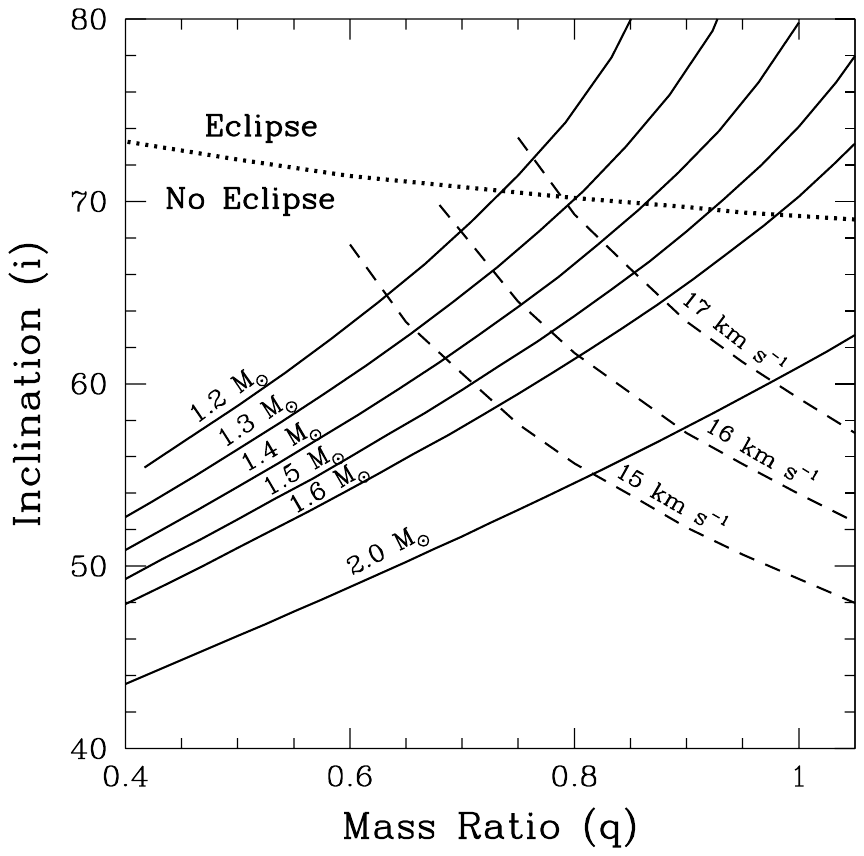}
\caption{
The inclination as a function of the mass ratio ($q$).  The solid lines are 
labelled with the  
NS mass.  The system does not eclipse, 
and the eclipse limit is shown by the dotted line.  The dashed curves show
the ellipsoidal relation between the $v\,sin\,i$ of the M~III and the 
mass ratio for the range of $v\,sin\,i$ values.  The intersection of 
the solid and dashed lines sets limits on $q$
and $i$. 
}
\label{fig:q_vs_inclination}
\end{figure}

The mass function, $\rm{f(m)}$, derived from our orbital parameters, 0.3337 $\pm$ 
0.0062 M$_\odot$ (Table \ref{table:orbit}), can be expressed as 
$$
\rm{f(m)}~=~ ( \, \rm{M}_\mathit{NS}^3~\mathit{sin}^3 \, \mathit{i} \,)/( \, \rm{M}_\mathit{RG} \, + \, \rm{M}_\mathit{NS} \, )^2.
$$ 
This can be re-written with the use of the mass ratio $q$ as 
$$
\rm{f(m)}~=~( \, \rm{M}_\mathit{NS}~\mathit{sin}^3 \, \mathit{i} \,)/( \, 1~+~\mathit{q} \, )^2.
$$
The relation between $q$ and $i$ from the mass function is shown in Figure \ref{fig:q_vs_inclination}. 
Curves are shown spanning the range of NS masses, 1.2 -- 2.0 M$_\odot$.


For Roche lobe filling systems the stellar radius, $R$, is a function of $q$ 
and the semi-major axis, $a$. The formulae are summarized in 
Table 2.1 of \citet{warner_1995}.  Since the NS has a very
small radius compared to the giant, the eclipse angle is arctan(R/a) 
and can be expressed as a function of q \citep{belczynski_mikolajewska_1998}.
The limiting angle for an eclipse as a function of $q$ is shown in
Figure \ref{fig:q_vs_inclination}. IGR~J16194$-$2810 does not eclipse,
and this sets a limit for the inclination as a function of $q$.


The rotational broadening of the lines is related to the rotational velocity, $v$, the
inclination $i$, the stellar radius, R, and the rotation period, P,
by
$$
v \, sin \, i ~ \lesssim ~ ( \, 2 \pi \, \rm{R} \, sin \, i \, ) / \rm{P}
$$
The M2 III fills its Roche lobe and the radius is by definition
the Roche radius.  Similarly, the rotation period and orbital period are 
equal \citep{morris_1985}.  With Kepler's third law, the Roche lobe filling stellar radius can be expressed
as a function of $q$ and orbital period.  This results in a relation between $q$ and $sin~i$,
also plotted on Figure \ref{fig:q_vs_inclination}.
The intersection of the relations from the mass function 
and $v\,sin\,i$ gives $q$ and $i$.
For the mean value of $ v\,sin\,i~=$ 16 km s$^{-1}$, $q$ is in the range 0.7 -- 0.9.  Over the entire 
range of possible $v\,sin\,i$, $q$ = 0.64 -- 0.98. The inclination 
is the range $i$ $\sim$ 55 -- 70$^\circ$. 

The filling factor, $f$, for Roche lobes in ellipsoidal systems
can be computed from $q$ and $i$ 
\citep{nie_et_al_2012}. 
Employing the above range of values for $q$ and $i$ and taking $\Delta R ~ = ~ \Delta I / 0.87$,
where $\Delta R$ and $\Delta I$ are the 
amplitudes of the light curve in the $R$ and $I$ bands and $\Delta I~=~0.11$,
the filling factor $f$ for the M2 III in IGR~J16194$-$2810 is   
in the range 0.90 -- 0.99.  This range 
for $f$ and related analysis by \citet{nie_et_al_2012}
confirms that 
the Roche lobe is filled.  

\subsection{M Giant Mass and Core Mass}\label{sec:mass_grav}

\begin{figure}
\centering
\includegraphics[width=.8\linewidth]{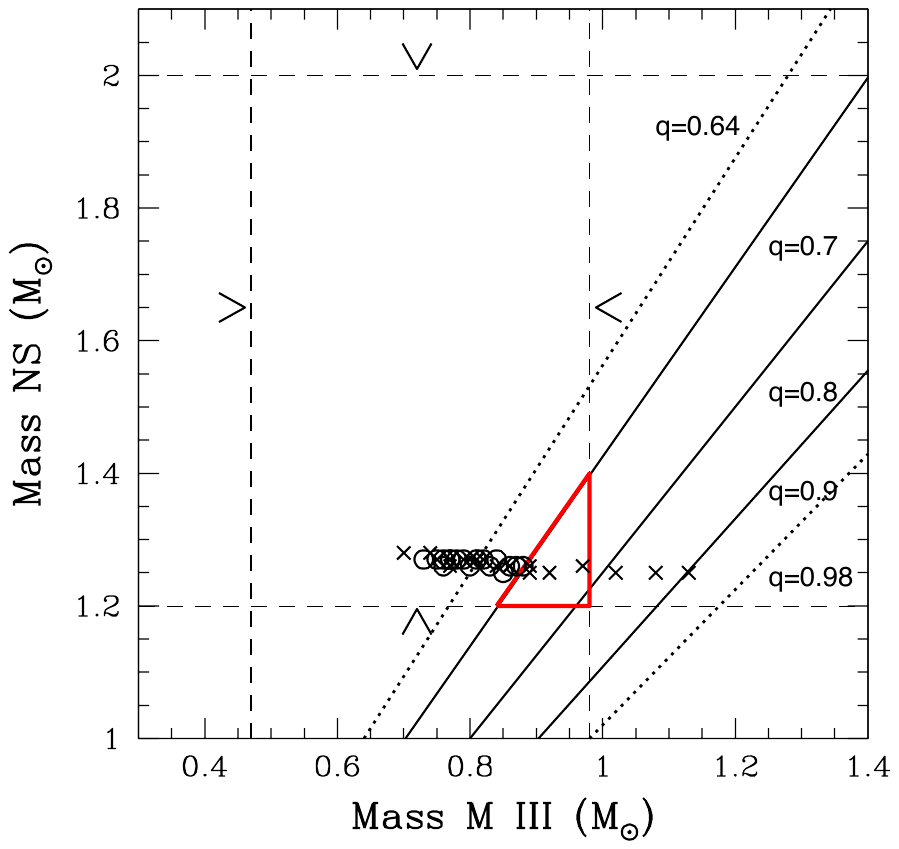}
\caption{
The range of the NS mass is shown as a function of the derived 
M giant mass and derived values of the mass ratio $q$.  The dashed 
lines and carets show the limits on the NS mass, 1.2 -- 2.0 M$_\odot$ 
and the M III mass 0.47 -- 0.98 M$_\odot$. The solid 
curves are values of $q$ for a projected rotation velocity of 
16 km s$^{-1}$.  The dotted lines are extreme values, $q$ = 0.64 and 0.98.
The masses
of the two binary components fall in the triangle,
NS mass 1.2 -- 1.5 M$_\odot$ 
and M III mass 0.75 -- 0.98 M$_\odot$ with the red triangle 
corresponding to more probable values of q. X symbols locate the model 
M$_2$ (M III) versus M$_1$ (NS) at log(L$_2$)=2.76, open circles
at orbital period 192 days.
}
\label{fig:mns_versus_mgiant}
\end{figure}

Since the M giant fills its Roche lobe, the radius of the Roche lobe is the radius of the M giant.  Defining
$ \Omega_b ~  = ~ 2\pi/P_b $, where $P_b$ is the binary period,
$$
{ { G M_{RG} } \over  {R_L^3} } = 10 \Omega_b^2
$$ 
where G is the gravitational constant \citep{phinney_kulkarni_1994}.  
From the luminosity and effective temperature, the radius of the 
red giant is 58 $\pm$ 7 R$_\odot$ (Section \ref{sec:temp_lum}).  
With this value and a rotation period of 192.5 days, the mass of the 
M giant is 0.70$^{+0.28}_{-0.23}$.  The M giant to NS mass ratio can
be found from Figure \ref{fig:q_vs_inclination}.  The extreme range 
of $q$ is 0.64 -- 0.98.  For the mean projected rotational velocity, 
16 km s$^{-1}$, the range is $q$ = 0.7 -- 0.9.  The range of NS masses 
resulting from $q$ and the M giant mass are shown in 
Figure \ref{fig:mns_versus_mgiant}. The NS lower mass limit of 
1.2 M$_\odot$ combined with $q$ limits the lower mass of the M giant to 
$\gtrsim$ 0.8 M$_\odot$.   Similarly, the upper mass limit for the M giant 
(Figure \ref{fig:mns_versus_mgiant}) limits the maximum mass of the NS to 
$\lesssim$1.5 M$_\odot$.  
The resulting value for the M III mass is 0.91$\pm$0.07 M$_\odot$  and for the NS 1.35$\pm$0.15 M$_\odot$.
For the 58 $\pm$ 7 R$_\odot$ M III radius, the 
volume average surface gravity of the M2 III is 
$\sim$ 5 -- 11 cm s$^{-2}$, i.e. $\rm{log(g)} \sim 1$.

The luminosity of a red giant branch (RGB) star depends only on the 
mass of the degenerate core, as shown by the  Paczy\'{n}ski relation 
\citep{paczynski_1970, tuchman_et_al_1983}.   For giants of solar 
metallicity, 
$$
L ~ = ~ (M_c/0.16 )^8 .  
$$
where the luminosity, L, is in L$_\odot$ and M$_c$ is the 
degenerate helium core mass 
in M$_\odot$\citep{phinney_kulkarni_1994}.  For IGR~J16194$-$2810, 
L$~=~573^{+72}_{-79}$ L$_\odot$ so $M_c$ is 0.35$\pm$0.01 M$_\odot$.  
Ellipsoidal variables have a period -- luminosity relation so there
must also be a relation between the orbital period and the core mass.
\citet{tauris_savonije_1999} provide a fit to models relating the 
core mass to the orbital period.  From their equation 20 with 
population I constants, $M_c$ is 0.35 M$_\odot$.
The \citet{tauris_savonije_1999} models are in thermal equilibrium.
While Roche-lobe filling, IGR~J16194$-$2810 currently does not have an extreme mass loss rate 
(Section \ref{subsec:accretion}), with most of the envelope mass of the M III intact 
(Section \ref{subsec:ellipsoidal_evol}), and this criteria should apply.  The agreement of the Paczy\'{n}ski and 
Tauris-Savonije values supports this.

\citet{renzini_et_al_1992} found that since the stellar envelope is convective,
the radius of the giant is also a function of the degenerate core mass.
From \citet{phinney_kulkarni_1994}  
$$
R ~ \simeq ~ 1.3 \left( \frac{M_c}{0.16} \right) ^ 5 
$$  
where the radius, R, is in R$_\odot$.
Thus the stellar radius and, in RLO, the size of the Roche lobe,
can be determined from only the luminosity, 
albeit with considerable uncertainty.  Alternately, the core mass
can be found from the radius.  The resulting core mass is
0.34$\pm$0.1 M$_\odot$, consistent with values found above.

\section{ABUNDANCES}\label{sec:abund}

Information on the age and evolutionary history of the system potentially can be 
determined from the star's chemical composition.  In addition, in several 
places in the above discussion, constants have been used appropriate for stars
of near-solar metalicity and this should be confirmed.
To determine the [Fe/H] 
abundance, we identified Fe I lines that were not significantly blended 
with other absorption features, located in spectral regions where the 
continuum was relatively well defined, and were included in the 
1.51 -- 1.70 $\mu$m region of the $H$-band covered by the 
$APOGEE$\footnote{Apache Point Observatory Galactic Evolution Experiment} line
list \citep{shetrone_et_al_2015}. 
Synthetic spectra were used to measure abundances and identify blends 
with the MOOG package \citep[][Version 2019]{sneden_1973} 
and MARCS\footnote{Model Atmospheres with a Radiative and Convective Scheme} 
model atmospheres \citep{gustafsson_et_al_2008}. 
As a starting value, the solar abundances of 
\citet{asplund_et_al_2009} were assumed. A microturbulence value of 
2.0 km s$^{-1}$ 
was used, a typical value for early M 
giants \citep{smith_lambert_1985,maas_pilachowski_2021}.

Grids of synthetic spectra were created to measure [Fe/H] abundances 
via $\chi^2$ minimization.  Atmospheric models were iterated to match the 
newly derived [Fe/H]. Similar analysis of the Arcturus spectrum 
\citep{hinkle_et_al_1995} produced an abundance consistent with the 
results of \citet{ramirez_allende_prieto_2011}. The 
standard deviation calculated from the abundances derived for individual 
Fe I lines was adopted as the statistical [Fe/H] uncertainty. Models 
at $\pm$100 K in T$_{eff}$ and $\pm$0.20 in log(g) were created individually, the new FeI abundance
computed, and the difference adopted as the uncertainty. The atmospheric parameter 
uncertainties were combined in quadrature with the statistical uncertainty and the total 
uncertainty is quoted for the abundance, [Fe/H] = $- 0.14 \pm 0.12$


We also sought to determine the 
C, N, and O abundances using selected spectral lines from the 
near-IR transitions of CO, CN, and OH.  
The line lists used were \citet{li_et_al_2015} for CO,
\citet{brooke_et_al_2014} for CN, and \citet{brooke_et_al_2016} for OH.
Lines from the $H$-band CO second overtone could not be identified.  
For CO we selected
a list of least blended lines in the $K$-band first overtone by
using the technique discussed in \citet{hinkle_et_al_2016}.
As we did for the [Fe/H] analysis,
abundances were derived with MOOG spectral synthesis software
and MARCS 1D atmospheric models.  A MARCS model was selected 
matching 
T$_{eff}$ = 3790 K, log(g) = 0.8, [Fe/H] = $-$0.14, and
microturbulence = 2 km s$^{-1}$.
The abundance results are summarized in Table \ref{table:abundances}.  The CNO 
abundances sum up, within the uncertainties, to the solar CNO value 
scaled to the observed metallicity.

From synthesis of the first overtone CO, $^{12}$C/$^{13}$C\,=\,$22\pm3$.
The $^{12}$C/$^{13}$C ratio was also computed with the curve-of-growth 
method \citep{hinkle_et_al_2016}.  This technique yields 
$^{12}$C/$^{13}$C\,=\,30$^{+31}_{-16}$, in agreement with the synthesis 
result. Due to rotational broadening, the spectral lines are 
not as deep when observed at high resolution as typically observed 
in an M2 III.  Thus, the oxygen isotope lines could not be reliably measured.

The set of stronger CO lines in the IGR~J16194$-$2810 spectra
results in a higher abundance of [$^{12}$C/Fe], 0.27, than the 
weak lines.  The weak line abundance is used here.
The strong CO $\Delta$v=2 lines in red giants are formed in an
extended atmospheric region. This has previously been reported 
in single M giants by \citet{tsuji_2008}.


\section{DISCUSSION}\label{sec:discussion}

In this section the properties of the IGR~J16194$-$2810 system defined above 
are discussed and compared to modeled LMXB evolution.
The observed properties place IGR~J16194$-$2810 among 
several other classes of objects, notably, the SyXB and the sequence E ellipsoidal variables.
We will look briefly at these groups from a viewpoint of stellar evolution to
see if any additional information can be gleaned on IGR~J16194$-$2810.
Finally, a model specifically tailored to IGR~J16194$-$2810 is discussed.
We conclude with brief comments on the X-ray luminosity and mass transfer.

\subsection{Binary Evolution}

LMXBs terminate their evolution as NS \textendash~WD binaries.  The 
family of NS - WD binaries includes millisecond pulsars (MSP).  There are currently 
several hundred LMXBs known \citep{bahramian_degenaar_2023} 
and an equal number of MSP \citep{manchester_et_al_2005, ozel_freire_2016}. 
LMXB evolution can be separated into two parts, (1) the formation 
of a binary consisting of a NS in an orbit with a low-mass star, and 
(2) RLO evolution resulting in a NS \textendash~WD binary 
\citep{bhattacharya_van_den_heuvel_1991}. 
We briefly review the two stages in the next subsections.
IGR~J16194$-$2810 has completed stage (1) and is in stage (2).

\subsubsection{Formation of the NS -- M III}\label{subsec:prog_bin_evol}

The standard scenario to create a LMXB starts with a binary 
consisting of a primary star of mass $\gtrsim$ 8 M$_\odot$ and 
a low mass companion.  The primary evolves and expands to fill 
its Roche lobe. Radii of evolved massive stars are of the order several AU, 
larger than all but the most widely separated 
LMXB orbits.  The existence of LMXBs requires a CE phase during which the  
low mass companion spirals in as 
the envelope of the primary is ejected 
\citep{verbunt_1993}. The existence of a NS requires a SN. 
The CE phase reduces the mass of the primary 
before the SN occurs, allowing the system to stay bound 
\citep{van_den_heuvel_heise_1972}.

Following the CE stage, in
massive stars, 
M $\gtrsim$ 12 M$_\odot$, the star develops a degenerate iron core that 
exceeds the Chandrasekhar limit and collapses. 
NS formed from Fe-core collapse are associated with large
kick velocities, typically $>$200 km s$^{-1}$ \citep{hobbs_et_al_2005}.
For
a narrow range of primary masses, $\sim$11 M$_\odot$, the 
stripped core becomes a helium star.  The helium star will undergo an 
electron-capture supernova with 
minimum loss of mass and a small kick velocity 
\citep{siess_lebreuilly_2018, yungelson_et_al_2019}.
In the mass range 8 -- 10 M$_\odot$, the star undergoes an AGB stage and
terminates its nuclear burning lifetime as an
oxygen-neon white dwarf (ONe WD) of high mass 
\citep{podsiadlowski_et_al_2004, limongi_et_al_2024, smartt_2009}.
The massive WD   
accretes mass as the companion star undergoes RLO 
and then undergoes accretion induced collapse (AIC),
to produce the NS \citep{kalogera_webbink_1998}. 
The first two processes produce a NS with an 
age only a few Myrs less than that of the binary system.  In
AIC the NS is born only when the companion evolves to fill its Roche lobe. 

Using precision orbital and NS spin properties that can be derived 
for binaries containing a pulsar, information on the 
NS progenitor can be derived \citep{knigge_et_al_2011}.  However, 
deriving similar information on the NS progenitor in LMXBs that 
do not contain a pulsar is
not possible.  The orbits are known to much lower precision and 
the time since the SN has erased information even in the most favorable cases.  
For example,
models (Section \ref{subsec:funs_models}) show that the elapse time between 
an AIC event and the start of the RLO   
is $\gtrsim$ 10$^5$ yrs.  
In the case of IGR~J16194$-$2810
there is no associated cataloged remnant and the field of IGR~J16194$-$2810, 
examined in near-infrared and visual images\footnote{http://cdsportal.u-strasbg.fr/  
The Two Micron All Sky Survey (2MASS) $K$ image 
shows diffuse emission extending approximately 2 arcminutes north and west.  
However, this is not seen in 
any other images, including AllWISE.}
shows no conspicuous nebulosity. 

Similarly, as a result of a SN explosion, the resulting NS and, if 
present, binary companion(s), will have a kick velocity relative 
to the field stars. The rotation of the immediate SN progenitor \citep{pfahl_et_al_2002}, the nature of the SN, 
the details of the 
pre-SN  binary orbit, and the symmetry of the SN explosion all play 
a role in setting the kick velocity \citep{suarez-andres_et_al_2015, 
stevenson_et_al_2022}.  A notable feature of IGR~J16194$-$2810 is 
that its space velocity, $-$57.7 km s$^{-1}$, is similar to that 
of other stars in that region of the Milky Way, i.e. 
IGR~J16194$-$2810 has a low kick velocity.  Other long period 
LMXBs are known with low kick velocities 
\citep{hinkle_et_al_2019,hinkle_et_al_2020} but since multiple 
factors determine the kick, this, like the
absence of a remnant, does not provide information on the progenitor of
an individual NS.

The M giant in IGR~J16194$-$2810 was shown to have very slightly metal poor abundances 
with no unusual abundances in Fe, C, N, or O that could be 
attributable to unusual evolution or near presence to a SN.
The abundances, [C/Fe] = $-0.19 \pm 0.11$, [N/Fe] = $0.31 \pm 0.11$, 
and [O/Fe] = $0.14 \pm 0.13$, are typical for field first ascent giants 
\citep{lambert_ries_1981, smith_lambert_1985}.  These values, as 
well as $^{12}$C/$^{13}$C = $22 \pm 3$, are all in agreement with material 
processed in the stellar interior that is being mixed to the surface 
during the first dredge up with standard mixing \citep{halabi_eid_2015}. 
The [Fe/H] = $-0.14 \pm 0.12$, similarly, is a typical value for a 
field giant \citep{anders_et_al_2014}.

The pre-SN CE event that ejects the massive star envelope
does not lead to significant mass loss from the low mass star
\citep{wang_et_al_2022}. 
The amount of ejecta captured by a companion, $\Delta m$, resulting 
from the SN can be estimated as 
$$
\Delta m~\sim~\Delta M_1 \left( \frac{R_2^2}{4 a_o^2} \right) f
$$ 
where $\Delta M_1$ is the mass lost from the massive star at the time of the SN, $R_2$ the radius of the
companion, $a_o$ the semimajor axis at the time of the SN explosion, and $f$ is an efficiency factor
\citep{suarez-andres_et_al_2015}.  
Assuming $f=1$, for IGR~J16194$-$2810 the mass of ejecta captured will be on 
the order of 0.01\% of the total SN ejecta.
The current mass of the M III envelope can be estimated from 
the mass of the M III (Table \ref{table:parameters})
of 0.84 -- 0.98 M$_\odot$ less the core mass of $\sim$0.35, i.e. $\sim$0.5 -- 0.6 M$_\odot$.
At the time of the first dredge up the envelope mass was likely $\sim$1 M$_\odot$.
Even in the event that the 
SN ejecta is several solar masses, the ejecta impacting the low mass companion 
is small compared to the mass of the 
stellar envelope. 
The envelope is convective and this material is mixed and ultimately makes no measurable 
difference in the abundances.
For other 
SyXB systems discussed in 
this series of papers \citep{hinkle_et_al_2006,hinkle_et_al_2019,hinkle_et_al_2020},
no abundance peculiarities have been measured.
Among shorter period LMXBs there are systems with detectable 
abundance abnormalities related to the SN \citep{gonzales_hernandez_et_al_2005,suarez-andres_et_al_2015},
however, these stars are in much closer orbits and do not have convective envelopes.

\subsubsection{Ellipsoidal Evolution}\label{subsec:ellipsoidal_evol}

The nuclear evolution of a red giant core is independent of
conditions in the convective envelope \citep{wood_faulkner_1986}. 
Once a star of increasing luminosity
fills its Roche lobe, the transfer of the stellar envelope of the donor star continues
as long as the luminosity of the donor increases.  
From the relations given in  Section \ref{sec:mass_grav},
it follows that the final binary orbital period, P$_b$, in days is 
$$
P_b ~ \sim ~  1.3 \left( \frac{M_{fc}}{0.16} \right) ^ 7 
$$
\citep{phinney_kulkarni_1994}
where the final core mass $M_{fc}$ is in solar masses. 
The final core mass for IGR~J16194$-$2810 will clearly be larger than 
the current core mass, so $\gtrsim 0.35$ 
(Section \ref{sec:mass_grav}).  Thus, P$_b$ will be $\gtrsim 310$ days
with considerable uncertainty due to the 7$^{th}$ power of the core mass. 
\citet{smedley_et_al_2014} provides an exponential expression that
results in a similar period but, again, with very large uncertainty due to the uncertainty in the core 
mass. 

Evolutionary tracks of LMXBs computed by \citet{tauris_savonije_1999} 
include some with roughly similar parameters to the 
IGR~J16194$-$2810 system.  Among these is a simulation with 
an M III of initial mass 1 M$_\odot$,
L $=$ 368 L$_\odot$, M$_c$=0.336 M$_\odot$, 
initial orbital period of 60 days, and a 1.3 M$_\odot$ NS. While a 
factor of two less luminous than IGR~J16194$-$2810 with a slightly 
lower core mass and shorter orbital period, this model is 
representative of the family of longer orbital period NS - M III binaries.  
The model has a mass transfer period of 13 Myr with an initial 
mass-transfer rate to the NS of $\sim$ 6 times the Eddington 
accretion rate, which drops to about twice the Eddington rate 
at the end of the mass transfer phase.  The Eddington accretion 
rate of a 1.3 M$_\odot$ NS for hydrogen-rich matter is 
$\sim$1$\times10^{-8}$ M$_\odot$ yr$^{-1}$ \citep{tauris_et_al_2012}.  The NS accretes 0.20 
M$_\odot$ over the 13 Myr mass transfer period. A similar simulation 
by \citet{podsiadlowski_et_al_2002} for an initial 1 M$_\odot$ giant 
with a 0.341 M$_\odot$ core results in a 0.43 M$_\odot$ He WD 
in a 617 day orbit.  The initial and final NS mass were 1.4 M$_\odot$ 
and 1.48 M$_\odot$.  While the average mass transfer rate is similar, 
a much larger peak rate is indicated with a shorter total duration 
of the mass transfer period, 4.5 Myr.

The IGR~J16194$-$2810 M III has a core mass of $\sim$0.35 M$_\odot$ 
and a total mass of 0.84 -- 0.98 M$_\odot$, so the envelope mass of 
IGR~J16194$-$2810 is currently about 60\% of the total 
donor mass.  A typical value for similar systems at the start of RLO 
is 70\% \citep{tauris_van_den_heuvel_2006}.  This implies that the 
IGR~J16194$-$2810 mass-transfer period is near the beginning with the 
NS mass near its original value.  Masses measured for the NS in 
binary NS -- He WD systems are $\sim$1.5 -- 1.6 M$_\odot$ 
(Section \ref{mass_ns}) and obviously include the accreted mass.  
This suggests a current NS mass near 1.3 M$_\odot$.


\subsection{Comparisons}

IGR~J16194$-$2810 has variously been classified as an  ellipsoidal binary,
a SyXB, and a LMXB. LMXBs are progenitors of  NS -- He WD.
Here we compare IGR~J16194$-$2810 to these groups.

\subsubsection{Ellipsoidal Variables}\label{subsec:ellipsoids}

\begin{figure}
\centering
\includegraphics[width=0.8\linewidth]{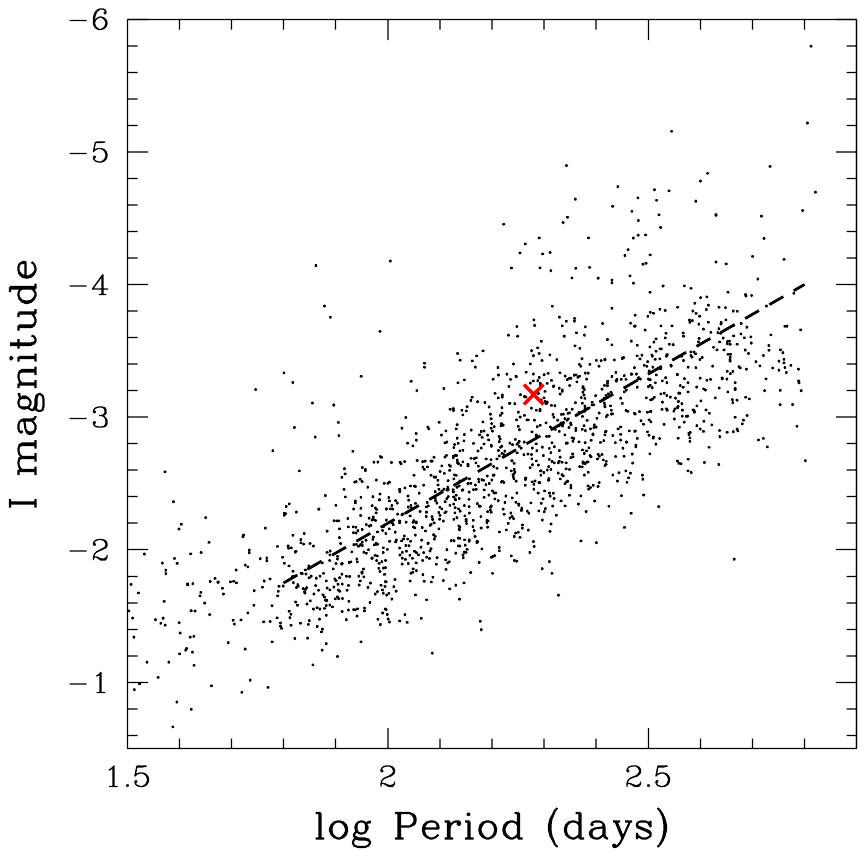}
\caption{
The Ellipsoid Period-Luminosity (P--L) diagram.  The dots are 
ellipsoidal variables in the LMC \citep{soszynski_et_al_2004} with an
adopted distance modulus of 18.5. No de-reddening has been applied.
The dashed line is the fit in the interval $1.75~<~log(P)~<~2.85$ from 
\citet{nie_et_al_2017}.  
Massive ellipsoidal variables are more luminous and appear at the 
upper edge of the P-L distribution.  Systems with M giants of mass 
$<1.85$ M$_\odot$ are scattered around the line \citep{nie_et_al_2017}.
The red $x$ marks the position of IGR~J16194$-$2810 for an observed 
$I~=~9.98$ mag and $A_I~=~1.54$ mag, i.e.~a de-reddened $I~=~8.44$ mag.  
Assuming a distance to IGR~J16194$-$2810 of 2101 pc, the absolute, de-reddened $I$ is $-3.17$ mag.  
}
\label{fig:Period_Lum}
\end{figure}

\citet{kiraga_2012} found IGR~J16194$-$2810 to be an ellipsoidal variable 
red giant. Ellipsoidal red giant binaries consist of a Roche lobe filling 
red giant with a typically unseen companion 
\citep{nicholls_wood_2012, pawlak_et_al_2014, nie_et_al_2017}.
The Roche lobe has a roughly ellipsoidal shape that results in brightness 
variations as the star rotates in the line of sight over the course 
of the orbit. 
\citet{zahn_1977} found that stellar rotation synchronizes with orbital 
period on time scales much shorter than the stellar evolution time scale.
In ellipsoidal variables, the Roche lobe rotates synchronously with the 
orbital period \citep{morris_1985}.  As a result, ellipsoidal variables   
form a period-luminosity sequence \citep[`sequence E' ][]{wood_et_al_1999}.  
In the case of IGR~J16194$-$2810 the ellipsoidal period and $I$ band 
magnitude, place it  
among the low mass stars on the ellipsoidal variable 
period-luminosity diagram (Figure \ref{fig:Period_Lum}).

In LMXBs the unseen companion is a NS while in field sequence E variables it is 
typically a main sequence star \citep{nie_et_al_2017}.  Thus the
mass ratio for the LMXB is reversed from sequence E variables, with the 
evolving star the less massive of the two.  While the companion in 
LMXB systems is not visually bright, it is hot and X-ray 
luminous.  The optical light curve has a bright spot resulting 
from heating of the M III by the NS X-ray flux.  However, as in the 
case of IGR~J16194$-$2810, light curve survey data is generally
too noisy to reveal subtleties.

The lifetimes of MS -- red giant ellipsoidal systems are 
less than 1 Myr, terminating when the red giant starts to 
overfill its Roche lobe resulting in a CE phase \citep{nie_et_al_2012}.
For low mass NS systems 
there is a similar rapid increase of mass exchange during the first Myr.
However, NS binaries are dynamically stable and super Eddington 
accretion onto the NS does not result
in a CE phase \citep{tauris_savonije_1999}.  As a result, lifetimes of 
NS -- ellipsoidal variables are much longer and limited 
by the mass exchange rate, the mass of the donor envelope, and the 
donor evolutionary status (Section \ref{subsec:funs_models}).


\subsubsection{SySt/SyXB}\label{subsec:SySt}

IGR~J16194$-$2810 is included on lists of SyXBs \citep{yungelson_et_al_2019}.
However, most SyXBs are not Roche lobe filling.  A typical 
system has an orbital period of years with the accretion onto the 
NS from the late-type giant/AGB wind \citep{hinkle_et_al_2016}.  
These are systems with multi-year orbital periods so that a RLO
phase, if it occurs, will occur on the AGB.

There are SySt with similar orbital periods to that of IGR~J16194$-$2810.
The shortest known orbital period for a well documented SySt is for the
WD$+$M4.5 III system T~CrB with a period of 227.5 days
\citep{fekel_et_al_2000}.  T~CrB is an ellipsoidal variable
and a recurrent nova
\citep{kraft_1958,belczynski_mikolajewska_1998}.
The more massive member of the T~CrB binary is a WD near the
Chandrasekhar limit \citep{selvelli_et_al_1992}. 
A possibly shorter period system is 
DASCH J075731.1+201735, 
an ellipsoidal M0 III with a possible WD companion, 
in a 119 day orbital period \citep{tang_et_al_2012}. 
Members of the SySt and SyXB classes are based on common 
observational characteristics but can have very different 
evolutionary paths.  While both T~CrB and DASCH J075731.1+201735 
are RLO systems, their evolutionary path is quite different and
they provide little additional insight into NS systems.


\subsubsection{LMXB}\label{subsec:LMXB} 

The majority of LMXBs have orbital periods of days.  From Kepler's third 
law, the corresponding semi-major axis is on the order of 0.04 AU, 
roughly 10 R$_\odot$, requiring a dwarf companion.   For orbital periods of 
a week the semi-major axis is near 0.1 AU, implying an evolved secondary.  
In all LMXB the donor star is Roche lobe filling with the 
brightest LMXBs having observed ellipsoidal variability.  For example,
the $V$=14.7 system Cyg X-2 is an ellipsoidal variable with a 9.8 day 
orbital period, a mass function of 0.69 $\pm$ 0.03 M$_\odot$, and 
mass ratio $q$ = 0.34 $\pm$ 0.04 with a A9III donor 
\citep{casares_charles_kuulkers_1998}.
\citet{orosz_kuulkers_1999} found the mass of the NS to be 1.78 
$\pm$ 0.23 M$_\odot$ and the mass of the secondary star to be 
0.60 $\pm$ 0.13 M$_\odot$, implying that the mass transfer is well 
advanced.

Less well known giant LMXB systems include
V395 Car = 2S 0921$-$630 with an orbital period of 9.0 days and  
accretion disk eclipses. The NS mass is 1.44 $\pm$ 0.10 M$_\odot$ with 
a companion mass of 0.35 $\pm$ 0.03 M$_\odot$
\citep{ashcraft_et_al_2012}.  The secondary has been identified as a 
low luminosity K0 III of $\sim$50 L$_\odot$ \citep{shahbaz_et_al_1999}.
The semi-major axis of the 2S 0921$-$630 donor is 0.1 AU.  This matches 
the radius of an early K giant, but an early K giant will be four 
times more luminous,   
placing the late-type star 
near the base of the giant branch.  
A yet more evolved system is GRO J1744$-$28 \citep{rappaport_joss_1997}.
In this 11.8 day orbital period system, the envelope of the progenitor 
solar mass G/K III has been almost entirely stripped 
leaving the donor with a mass $<$0.3 M$_\odot$ \citep{gosling_et_al_2007}.

The LMXB systems with the longest orbital periods, excluding 
IGR~J16194$-$2810, are GX 13+1 and GRS 1915+05 
\citep{bahramian_degenaar_2023}. 
GX 13+1 has an orbital period of 24.7 days and a K5III donor 
\citep{bandyopadhyay_et_al_1999}. 
GRS 1915+105 has an orbital period 33.9 days \citep{steeghs_et_al_2013} 
with a K III donor \citep{greiner_et_al_2001}.  However, this 
system hosts a BH primary.

The period distribution of LMXBs reflects the evolutionary time scale.
Short orbital period LMXBs contain a Roche lobe filling dwarf and 
require an evolutionary time scale of Gyr for the core to be stripped.  
For system containing a red giant this takes multiple Myr.  
As a result the number of LMXBs is heavily skewed to short orbital periods
and it is not surprising that few are known with longer periods.


\subsubsection{NS -- WD Binaries and Recycled Pulsars}\label{subsec:pulsars}

For approximately half of the binary pulsars, the mass of the secondary 
requires a He WD \citep{phinney_kulkarni_1994}.  
Modeling and observations of both the MSP and LMXB groups have secured 
the link between these groups \citep{smedley_et_al_2017}.
Mass transfer during the RLO spins up the NS and re-activates 
the pulsar \citep{phinney_kulkarni_1994}. 
About one-fifth of LMXBs have been found to be accreting 
millisecond X-ray pulsars. 
On the completion of the mass transfer these systems will turn into 
millisecond radio pulsars \citep{bahramian_degenaar_2023}.
Models by \citet{smedley_et_al_2014, smedley_et_al_2017} 
show that millisecond pulsars result from systems with initial donor masses 
in the range 0.875 --  1.20 M$_\odot$ with the pulsar spin period depending on the details of the mass 
accretion onto the NS during RLO.  

\begin{figure}
\centering
\includegraphics[width=0.9\linewidth]{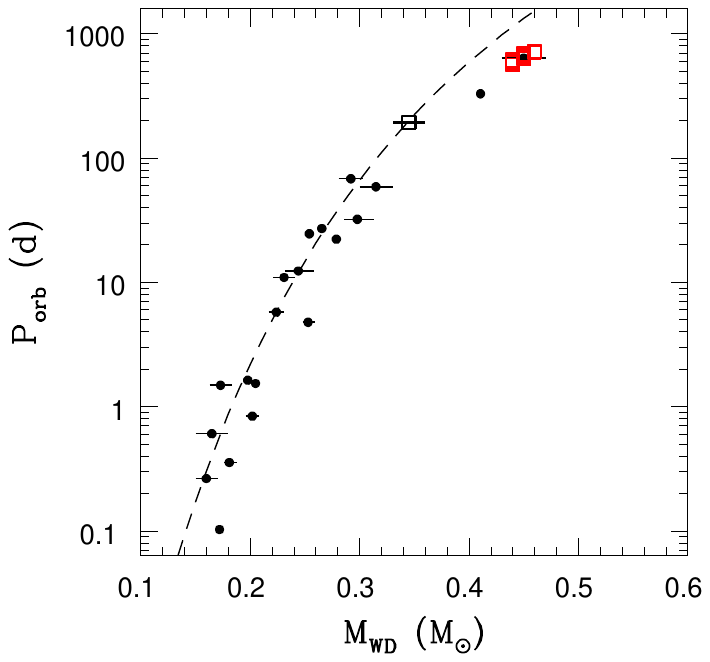}
\caption{
Orbital periods as a function of the WD masses 
for 20 NS -- He WD systems where the uncertainties in the WD mass
are no more than 0.02 M$_\odot$ \citep[Table 2 of][]{gao_li_2023}.  Where 
the uncertainties in the mass are large enough to be plotted they are shown as a horizontal line. 
The dashed line is the period -- white dwarf mass relation of \citet{lin_et_al_2011}.  
The measured period and the core mass, derived from the core-mass luminosity relation, is shown for 
IGR~J16194$-$2810 (open black square).  Predicted values for the evolutionary end product NS -- WD 
are taken from Table \ref{table:os_1} and shown as open red squares. 
}
\label{fig:p_orb_m_wd}
\end{figure}

Figure \ref{fig:p_orb_m_wd} shows the orbital 
period as a function of the core mass for the NS -- He WD systems 
with well established masses \citep{gao_li_2023}.  A relation 
between period and WD mass can be established by modeling systems 
\citep{lin_et_al_2011}, and this is also shown.
The current period and core mass of IGR~J16194$-$2810 are shown.
Since hydrogen shell burning during RLO increases the M III core mass, 
the core-mass orbital-period relation requires the future IGR~J16194$-$2810
NS -- WD system to have a longer period than the current orbital 
period of 192.5 days (Section \ref{subsec:ellipsoidal_evol}).   


Table \ref{table:pulsar_table} provides a list from the   
ATNF Pulsar Catalog of NS -- He WD pulsars with orbital periods 
longer than 200 days.  IGR~J16194$-$2810, at the long orbital 
period end of the LMXBs, clearly represents an unusual but not 
unique evolutionary path.   
Conclusions about the fate of IGR~J16194$-$2810 can be drawn by 
looking at the known NS -- WD systems.
Table \ref{table:pulsar_table} contains no fully recycled pulsars, 
i.e. pulsars with spin periods less than 10 ms.
Four systems contain a mildly (or close to mildly) recycled pulsar  
with spin periods $\lesssim$35 ms.
Three systems, J0823+0159, J1803$-$2712, and J1822$-$0848, have 
spin periods $>$ 300 ms.  Long orbital period systems like 
IGR~J16194$-$2810 are clearly not MSP progenitors.
The absence in Table \ref{table:pulsar_table} of  NS -- He WD pulsars 
with orbital periods longer than 1000 days suggests that 
IGR~J16194$-$2810 is at the long end of 
LMXB orbital periods for systems that can become NS -- He WD pulsars.  
There are very few long orbital period pulsars of any type.
One NS -- CO/ONeMg WD pulsar, J0823+0159, has an orbital period 
longer than 1000 days.


\subsection{FuNS Models for IGR~J16194-2810}\label{subsec:funs_models}

\begin{figure}
\centering
\includegraphics[width=0.9\linewidth]{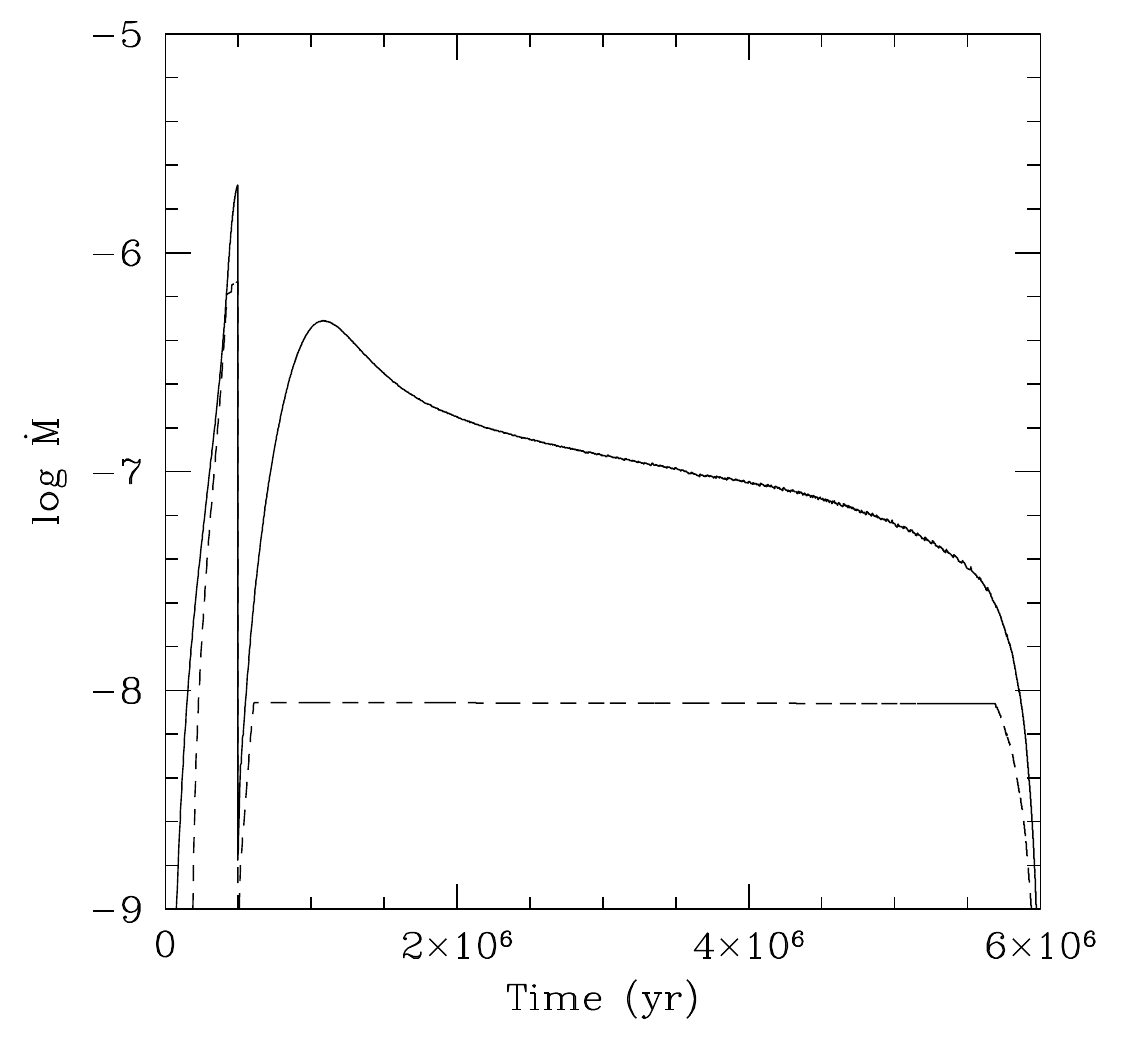}
\caption{
Mass transfer rate (solid line) and mass accretion rate onto the
WD/NS (dashed line) versus time, for the model G4 (see Table
\ref{table:os_1}). As described in the text, two distinct phases are
recognizable. Before the AIC, once the secondary RLO occurs mass transfer and accretion
to the massive WD is quite fast. After RLO of $\sim 5 \times 10^5$ yr, the
white dwarf attains the Chandrasekhar limit and collapses. Then,
for a short time, the system consisting of a bright giant and a neutron
star remains detached, until the RGB star again fills the Roche lobe. 
During this second mass transfer phase, that lasts for $\sim
5 \times 10^6$ yr, only a small fraction of the mass lost by the giant
star is actually accreted onto the NS surface. As a consequence of
the mass and angular momentum losses, the period and the orbital
separation progressively increase. The RLO ceases
when the mass of the giant is reduced to $\lesssim 0.5$ M$_\odot$.
}
\label{fig:os_1}
\end{figure}

In order to more fully investigate the evolution of IGR~J16194$-$2810 both as a current LMXB
and as the progenitor of a future NS -- WD binary,
FuNS\footnote{
FUll Network Stellar evolution code; \citet{straniero_et_al_2006}
}
code was configured for standard
interacting binary evolution. The FuNS calculation is very similar
to previous binary evolution calculations for the progenitors of millisecond pulsars with long orbital periods,
such as those reported by \citet{tauris_et_al_2013} and \citet{wang_et_al_2022}.
In particular, the calculation starts with a system consisting of a main sequence star (M2)
and a massive (ONe) white dwarf (M1). If the initial separation is large enough, the secondary will
fill the Roche lobe when it evolves on the red giant phase.
For the Roche lobe filling giant, the rate of mass-transfer through the internal Lagrangian point (L1) is calculated according 
to \citet{eggleton_2006}.  At the onset of RLO, the mass-loss rate suddenly increase up to a few 
time 10$^{-6}$ M$_\odot$ yr$^{-1}$ (Figure \ref{fig:os_1}).  The amount of mass actually accreted onto the WD is calculated 
by following the prescriptions of \citet{kato_hachisu_2004}. 
Since the total angular momentum is not conserved, due to 
mass lost by the system and emission of gravitational waves, variations of
orbital separation and period must be taken into account \citep{de_loore_doom_1992}.
Note, in particular, that according to a standard assumption, the fraction of the transferred mass, M$_{trans}$,
that is not accreted onto the compact companion, $\Delta M_{lost}\,=\,M_{trans}-M_{acc}$, is 
instantaneously lost by the system and that this mass removes from the system an amount of 
angular momentum equal to $J_{lost}\,=\,\Delta M_{lost} J_1$, where J$_1$ is the specific angular 
momentum of the WD or NS.


When the mass of the primary attains $1.38$ M$_\odot$, we assume that the WD undergoes an
induced collapse that leads to the formation of a neutron star. This occurrence causes a sudden loss
of gravitational binding energy,
so that the resulting neutron star mass is reduced to $\sim 1.25$ M$_\odot$
\citep[see, e.g., ][]{tauris_et_al_2013}. As a consequence, the orbital
separation suddenly increases and the system becomes temporarily detached.
We neglect the possible effect of a kick received by the newborn NS.
Later, the donor might fill the Roche lobe again initiating a second
mass-transfer phase.  
Now, the mass-loss rate is slower and comparable to, or lower than, the shell-H burning 
rate, so that the core mass and luminosity of the donor can possibly increase (Figure \ref{fig:os_1}).
The rate of the NS accretion is calculated as the minimum
between the mass transfer rate and the Eddington accretion limit, multiplied by a scaling factor
representing a sort of accretion efficiency. As discussed in,
e.g. \citet{tauris_et_al_2013}, there is increasing evidence of
inefficient accretion in LMXBs. As in \citet{wang_et_al_2022},
we tentatively assume a scaling factor for the NS accretion rate of $0.3$.

In order to test the code, we repeated some of the calculations discussed in \citet{wang_et_al_2022}
and we found a reasonable agreement.
In spite of this concordance, there are several uncertainties
in the theoretical description of the mass transfer process that
affect the quantitative result. These uncertainties concern, in
particular, the recipes used to calculate the mass transfer rate,
i.e.~the mass lost by the RGB star during RLO, the accretion rate
onto the compact companion, and the amount of mass and angular momentum
effectively lost by the binary system.

To reconstruct the evolution of a system
that reproduces the present-day observations, we have first selected
the most likely area in the
the initial parameter space. The hypothesis is that
the system is in a transient phase, when the companion
of the NS is evolving on the RGB. The giant is transferring
mass to the compact companion.
Hence,
the model must fit five observational constraints, C1 -- C5.
The present luminosity of the RGB
star is rather high, nearly 600 L$_\odot$ (C1), not too far from the RGB
tip, a fact that implies a large initial separation.  This is confirmed
by the large present-day orbital period, 192.5 days (C2). The present-day
mass, $\sim$1 M$_\odot$, is still high compared to its core mass, 0.35 M$_\odot$
according to the core mass-luminosity relation. Hence, a large
fraction of the envelope has not yet been lost, which implies that
a short time has elapsed since RLO was initiated,
no more than 10$^5$ yr (C3).

The initial configuration of the models is a main-sequence
star, with mass between 1.2 and 1.7 M$_\odot$ (M$_2$) and an ONe WD with mass between
1.2 and 1.3 M$_\odot$ (M$_1$). Then, M$_2$ evolves into a red giant until it fills
its Roche-Lobe (RL). 
The M$_2$ luminosity at
the beginning of the mass transfer phase depends on the initial
period (P$_i$) and, since the mass loss is so fast, it practically
does not change until the core collapse that creates the NS. In practice, during this first mass transfer
episode, the star progressively moves toward lower effective temperature and larger radii, while its 
luminosity does not change. 
textbf{
In any case, the M$_2$ luminosity
at the time of the AIC should be lower than the present-day luminosity,
that is: log L$_{AIC}$/L$_\odot<$2.8 (C4). The mass of the RGB star immediately
after the AIC depends on the initial M$_1$ and M$_2$ masses, but it is
practically independent of the initial period (i.e. separation). In
any case, it cannot be lower than the estimated present-day mass,
namely M$_2$ (post AIC) $\gtrsim$ 1 M$_\odot$ (C5).} After the AIC the system remains
detached for a short time.
As the luminosity increases, the
RGB star expands and fills the Roche lobe for the second time.  At this stage the
transfer rate is a few 10$^{-7}$ M$_\odot$ yr$^{-1}$ and the mass transfer timescale
is comparable to the shell-burning timescale, so that the star may
climb the RGB. Note that during this second RLO phase, the NS accretion occurs under a super-Eddington regime.
Meanwhile, the luminosity of the RGB companion increases, until its
envelope is almost fully removed. Then, the star contracts and turns
off the RGB. As a consequence, its radius becomes smaller than the
RL radius resulting in a detached configuration for the binary.


The results of our calculations are summarized in Table \ref{table:os_1} and
Table \ref{table:q_values}. There are three
sets of models, designated G, J, and H.  The initial M$_{1}$ is 1.30 M$_\odot$ for models G.  It is 1.25 M$_\odot$
and 1.2 M$_\odot$ for models J and H, respectively.

The initial configuration of the AIC models comes from a binary system
with a main sequence intermediate mass star ($\sim$ 9 M$_\odot$) and a
low mass star ($\sim$1.4 M$_\odot$).
After the C burning, the intermediate mass star enters a super-AGB phase
and fills its RL.  Taking the bolometric magnitude to be about $-$7
and the effective temperature about 3000 K then the super-AGB star
radius is of the order of 800 R$_\odot$, 3.7 AU.  By assuming a similar
RL radius, the separation of the binary should be of the order of
1200-1500 R$_\odot$, i.e. 5.6 -- 7.0 AU.  This results in a CE and all
the envelope mass of the intermediate mass star is lost, while the main-sequence
companion is marginally affected. The resulting binary will be a
ONe WD plus the original main sequence low-mass star with the initial
configuration shown for the models in Table \ref{table:os_1}.

\begin{figure}
\centering
\includegraphics[width=0.8\linewidth]{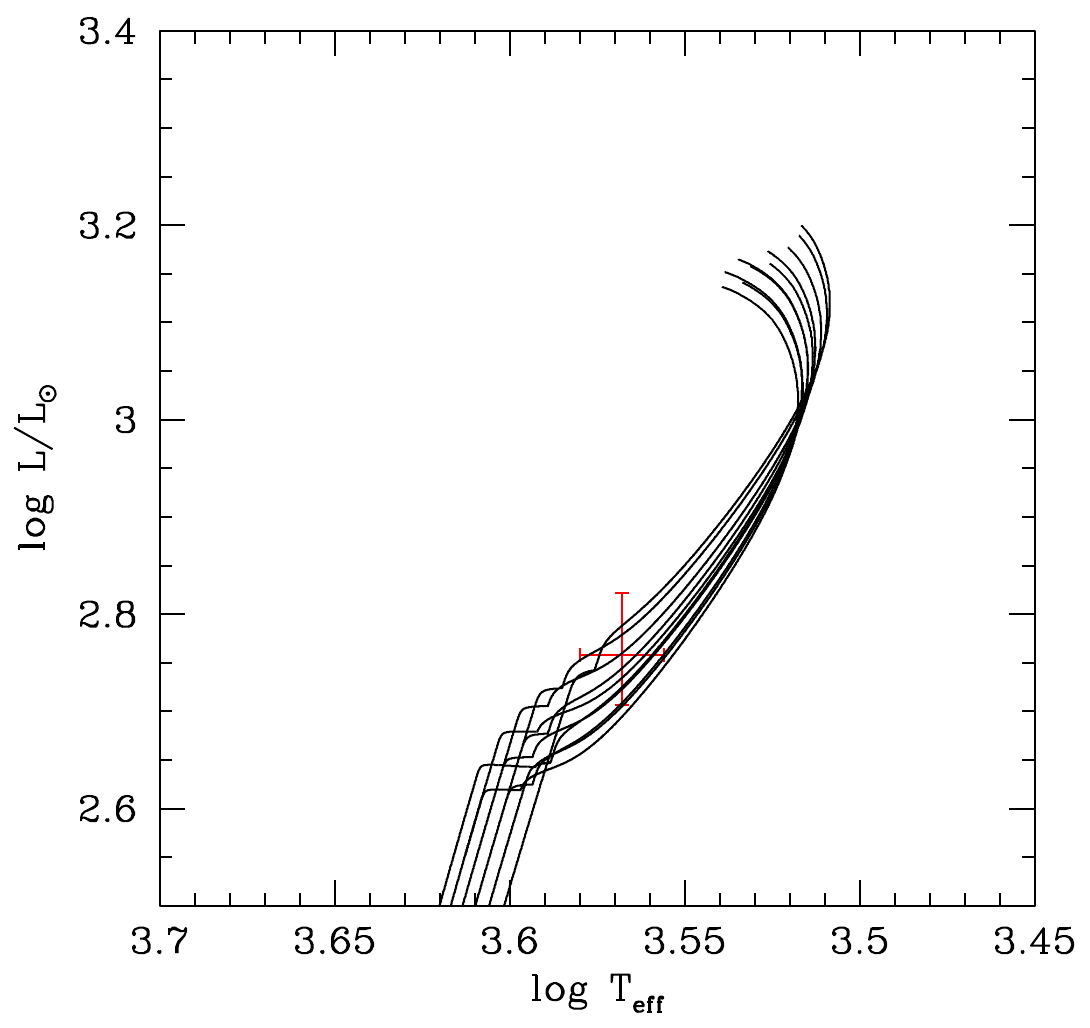}
\caption{
HR diagram of models G1 -- G9 (Table \ref{table:os_1}).  L-T$_{e\mkern0muf\!f}$ for IGR~J16194$-$2810 M$_2$ is shown (red) with uncertainties  
from Table \ref{table:parameters}.
}
\label{fig:os_2}
\end{figure}
\begin{figure}
\centering
\includegraphics[width=0.8\linewidth]{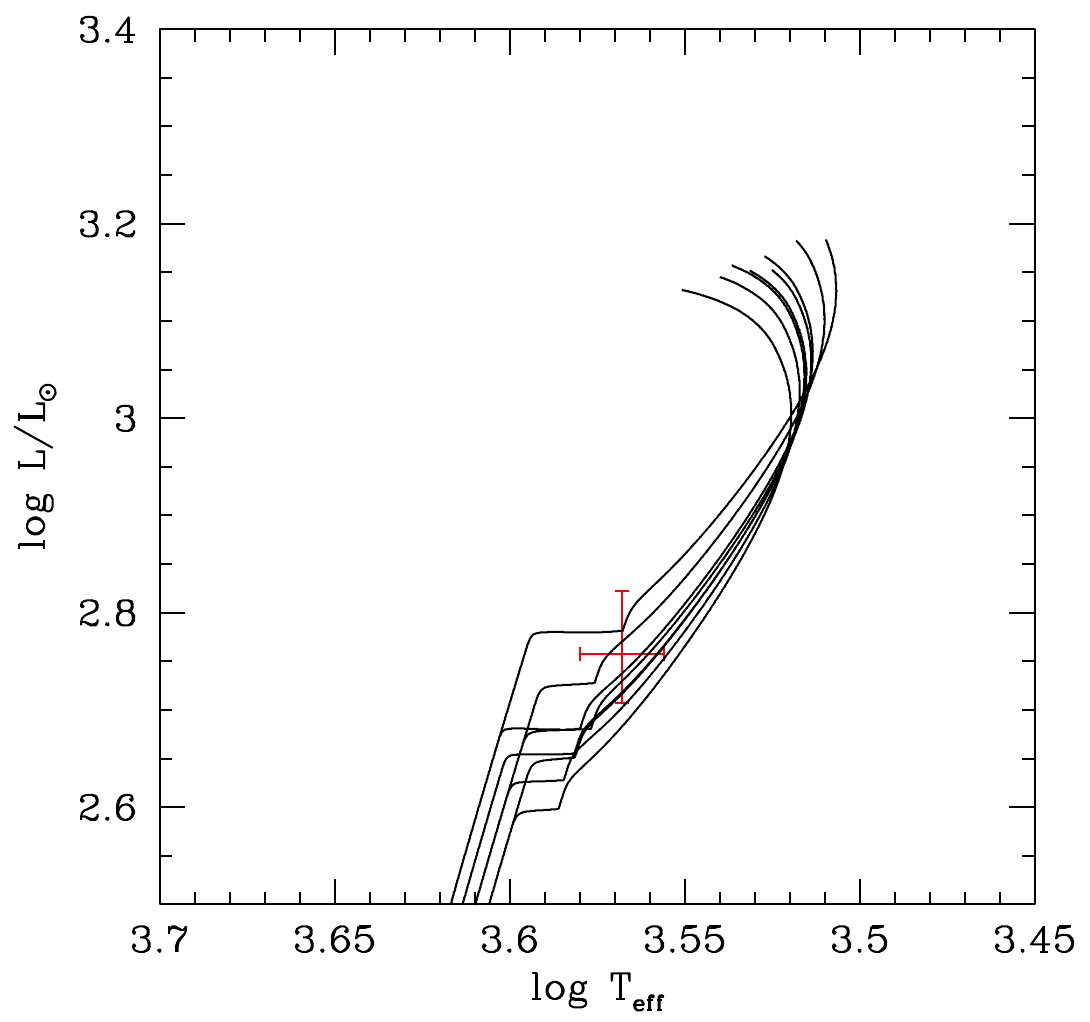}
\caption{
As in Figure \ref{fig:os_2} with models J1 -- J7 shown as black lines. 
}
\label{fig:os_3}
\end{figure}
\begin{figure}
\centering
\includegraphics[width=0.8\linewidth]{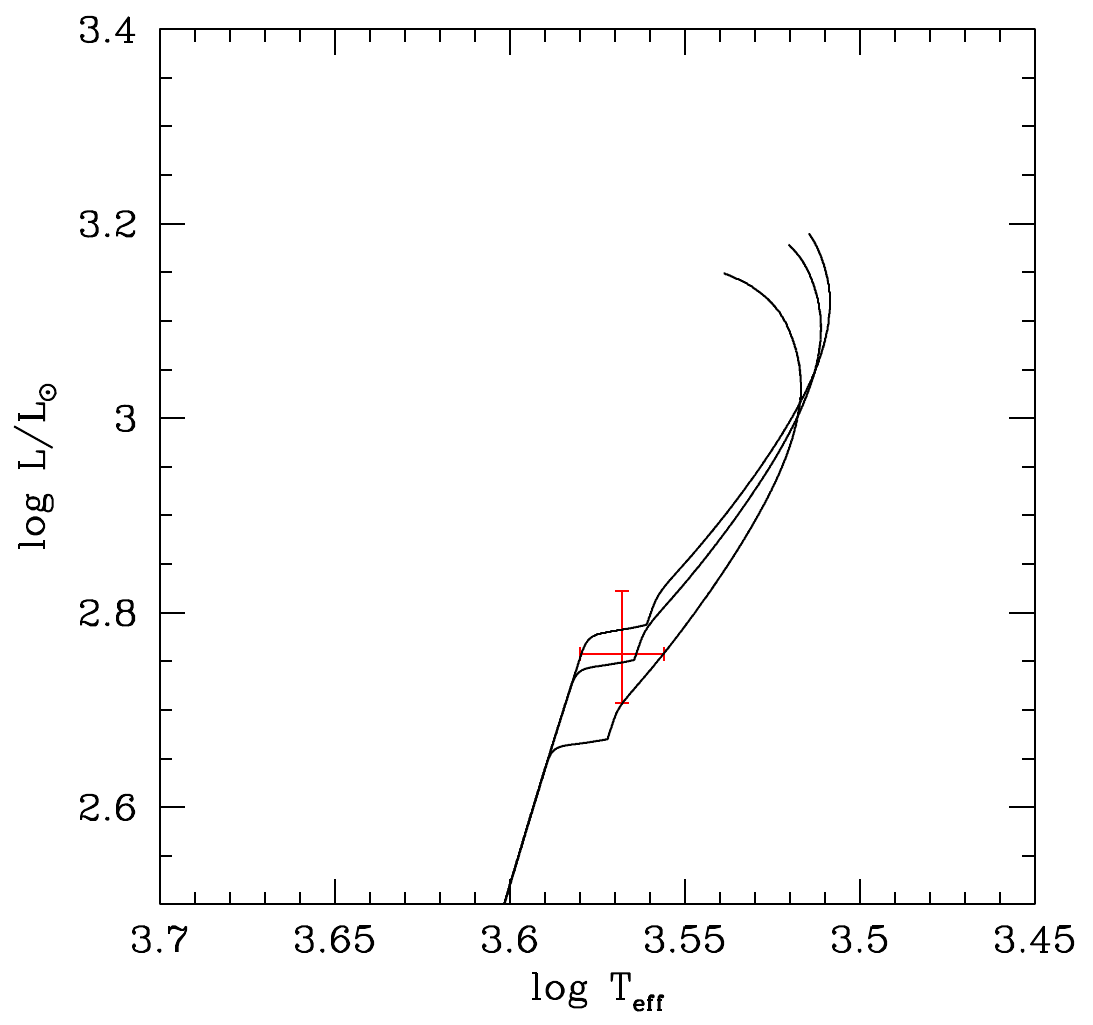}
\caption{
As in Figure \ref{fig:os_2} with models H1, H2 and H4 shown as black lines. 
}
\label{fig:os_4}
\end{figure}

HR diagrams of the three sets of models, G, J, and H, are presented in
Figures \ref{fig:os_2}, \ref{fig:os_3}, and \ref{fig:os_4}, respectively.
All the computed evolutionary tracks pass through the (L,T$_{eff}$) error box for 
IGR J16194-2810. In addition, the luminosity at which an orbital period P=192 days 
is attained is always within the estimated error bar.


The predicted NS mass is almost independent of the choice of the
initial parameter set (1.26 -- 1.27 M$_\odot$). The predicted mass of
the RGB star is often smaller than the measured value. Only in a few
cases is it within the uncertainties (Figure \ref{fig:mns_versus_mgiant}). In Table \ref{table:q_values}
the q values predicted at log L=2.76 and 
P=192 days, are shown for the G models.  At log L=2.76 for seven of the ten G models
the predict q ratio is within the estimated range (0.64 -- 0.98),
while when P=192 days, only four models are marginally compatible
with the empirical estimation (see a summary in Table \ref{table:q_values}).  A match/mismatch with observations
could be likely due to the many uncertainties affecting close-binary
evolution models, those related to the mass-loss rate, the
accretion rate, etc.  It should also be noted that these
models are made assuming that both the RGB star and the WD/NS
companion are non-rotating. Assuming that spins and orbit are
synchronized,  the rotation velocity of the RGB star is of the
order of 15 km s$^{-1}$, assuming a radius of 58 R$_\odot$ and P=192 days.
This is not a huge rotation velocity if the star is a rigid
rotator.  On the other hand, the spin up caused by
the AIC implies a quite fast rotation for the NS. This occurrence
could affect the accretion rate with possible implications for the
predicted NS mass and orbital separation/period.

All the models climb the RGB up to a luminosity close to the RGB tip. 
The final value of M$_2$, i.e. the mass of the secondary at the time of the 
binary system detachment, is always $\geq$ 0.44 M$_\odot$. Therefore, 
the possibility cannot be excluded that for some models the secondary 
will ignite He during the compression phase toward the WD cooling sequence. In 
that case, a late He-flash will occur that may trigger an ingestion of H by the 
flash-driven convective zone. Hence, the final outcome may be either a He WD or a CO WD. 
We will explore this late evolution scenario in a forthcoming paper.


\subsection{Mass Transfer and Accretion}\label{subsec:accretion}

X-ray luminosity (L$_X$)
can be converted
to  
a mass accretion rate by scaling by $c^2$ 
\citep{wiktorowicz_et_al_2017}.  
For 
IGR~J16194$-$2810,
L$_X$ is $2 \times 10^{35}$ ergs~s$^{-1}$ (Section \ref{sec:space_vel}),
corresponding to an accretion rate 
$\approx~3 \times 10^{-12}$ M$_\odot$ yr$^{-1}$. 
This value is order of magnitudes less than 
that expected for a RLO system, values close to 10$^{-8}$ M$_\odot$ yr$^{-1}$ are 
expected (Sections \ref{subsec:ellipsoidal_evol} and \ref{subsec:funs_models}, Figure \ref{fig:os_1}).
As noted by \citet{yungelson_et_al_2019},
the X-ray luminosity is similar to that expected
from M III wind accretion using 
Reimers' mass loss 
and
a mass transfer ratio, the mass accretion rate ratioed to the mass loss rate,
of $\sim$0.4\%, similar to values found in M giant -- WD symbiotics 
\citep{skopal_carikova_2015}.  

The NS spin period for IGR~J16194$-$2810 is 4.047 hours, 14570.34 s, 
detected in TESS photometry \citep{luna_2023}.
The NS spin period, orbital period, and X-ray luminosity 
of IGR~J16194$-$2810
fall among the wind accretion SyXB on L$_x$ -- P$_{orb}$ and the Corbet P$_{spin}$ -- P$_{orb}$ 
diagrams \citep{yungelson_et_al_2019}.
This is 
in contradiction to observations that 
IGR~J16194$-$2810 is a RLO LMXB as well as to 
models, for instance \citet{tauris_savonije_1999} and the FuNS models, that indicate a mass accretion rate  
in the super Eddington regime, typically about 0.35 of the Eddington rate, 
of about $1 \times 10^{-8}~\rm{M}_\odot$ yr$^{-1}$.  

One possible explanation is absorption of the X-ray flux by the circum-NS accretion disk.
\citet{koljonen_hovatta_2021} note that longer orbital periods in LMXBs translate
into larger Roche lobes and larger accretion disk sizes.
\citet{koljonen_tomsick_2020} discuss several systems with large accretion disks at high orbital
inclination that have intrinsic 
X-ray luminosities much lower than the Eddington luminosity as a result of heavy 
obscuration.  However, the examples cited, including the BH LMXBs V404 Cyg and GRS 1915+105, have
complex time dependent X-ray behavior, which is not seen in IGR~J16194$-$2810. 
On the other hand, varying obscuration by the accretion disk could be the cause of 
transient detections of the IGR~J16194$-$2810 spin period \citep{luna_2023}.


\section{CONCLUSIONS}\label{sec:conclusion}

A spectroscopic orbit has been determined for the M2~III 
star in the binary IGR~J16194$-$2810. This system was previously 
identified as an X-ray source and as an ellipsoidal variable. 
The M2~III is, as expected, in synchronous rotation with the orbital period.
In accord with the ellipsoidal variables, the photometric period is half of
the spectroscopic period.
The evolutionary track and stellar surface abundances show that 
IGR~J16194$-$2810 is a luminous first ascent giant with RLO.  
From values of the 
mass function and $v\,sin\,i$ derived from observations, the mass of the NS is shown 
to be in the range 1.2 -- 1.5 M$_\odot$ and the mass of the
M III 0.84 -- 0.98 M$_\odot$. 
From the luminosity, the core mass of the M III is 0.35 $\pm$ 
0.01 M$_\odot$. Hence, the envelope mass is large, $\sim$0.5 -- 0.6 M$_\odot$, 
and the system appears to be at an early stage in its evolution to become 
a NS -- WD binary.

The models show that the M III will climb the RGB up to a luminosity close to the RGB tip.
The mass transfer period will be about 5.5 Myr.
The final value of current M III will be $\geq$ 0.44 M$_\odot$. 
Some models indicate that the secondary
will ignite He during the compression phase toward the WD cooling sequence. In
that case, a late He-flash will occur triggering ingestion of H by the
flash-driven convective zone.  Hence, the WD companion to the NS 
could be either a He WD or a CO WD.
The models predict a final orbital period of $\sim$560 -- 720 days,
in agreement with the system evolving into the longest period 
group of NS -- WD pulsar binaries.   
The longest period NS -- WD binaries do not have MSPs, likely due to shorter mass 
transfer period than for lower luminosity donors.

IGR~J16194$-$2810 is a highly unusual variable star.  
It is an atypical member of the class of ellipsoidal variables
where most members are MS -- late-type giant binaries. 
The characteristics of IGR~J16194$-$2810 overlap various classes 
of mass exchange giant binaries, most notably the symbiotics.  
It is included in lists of SyXB systems but most of the other members of this class have 
much longer orbital periods. 
From both a classification and evolutionary perspective it 
is a RLO LMXB but, again atypical, with the longest period,
by a factor of more than three, of these objects. The majority of
the LMXBs are systems with a star evolving off the MS.  IGR~J16194$-$2810
is at the opposite end of the luminosity distribution.  Stars that are yet 
more luminous can not complete the removal of the donor envelope on the
giant branch and will go on to become AGB systems. 
The M III in 
IGR~J16194$-$2810 is a relatively bright star and presents an interesting object in which
to observe the LMXB process.


\acknowledgments 
We thank an anonymous referee for highly insightful comments
made from a very different perspective on the evolution of the binary than that  
presented in the original draft of this paper. 
This resulted in significant improvement in the discussion section.
We thank a second referee for a detailed review that resulted in 
a refinement of the input parameters.
We thank Luciano Piersanti for helpful discussion about binary evolution models
and Bernhard Aringer for useful discussions.  NOIRLab reference 
librarian Sharon Hunt helped us access some much needed reference materials.  
SM plot, developed by Robert Lupton and Patricia Monger, was used 
in the production of some figures. This research was facilitated by 
the SIMBAD database, operated by CDS in Strasbourg, France, the 
VizieR catalogue access tool, CDS, Strasbourg, France (DOI: 10.26093/cds/vizier),
and the Astrophysics Data System Abstract Service, operated by 
the Smithsonian Astrophysical Observatory under NASA Cooperative 
Agreement NNX16AC86A.  Spectrum synthesis was carried out using MOOG. 
The MOOG collaboration and grant support 
from the U.S. National Science Foundation are acknowledged.
This work made use of data from the 
European Space Agency (ESA) mission Gaia 
(\url{https://www.cosmos.esa.int/gaia}), processed by the Gaia Data 
Processing and Analysis Consortium (DPAC). Funding for the DPAC 
has been provided by national institutions, in particular the institutions 
participating in the Gaia Multilateral Agreement.  The Phoenix spectrograph,
used in this research, was 
developed at NOAO.  This work used IGRINS,
developed under a collaboration between the University of Texas at Austin and the Korea Astronomy and Space 
Science Institute (KASI) with the financial support of the US National Science Foundation under 
grants AST-1229522 and AST-1702267, of the University of Texas at Austin, and of the Korean GMT Project of KASI. 
ZGM is partially supported by a NASA ROSES-2020 Exoplanet Research 
Program Grant (20-XRP20 2-0125).  
Astronomy at Tennessee State University was supported  by the 
State of Tennessee through its Centers of Excellence program. 
KHH and RRJ 
thank the NOAO Office of Science and the NOIRLab RSS group for support of their research.  
NOIRLab is managed by the Association of Universities for Research in Astronomy (AURA) under a 
cooperative agreement with the National Science Foundation.

OCHID identification numbers:

FRANCIS C. FEKEL [0000-0002-9413-3896] 

KENNETH H. HINKLE [0000-0002-2726-4247] 

RICHARD R. JOYCE [0000-0003-0201-5241]

THOMAS LEBZELTER [0000-0002-0702-7551]

ZACHARY G. MAAS [0000-0002-0475-3662]

MATTHEW MUTERSPAUGH [0000-0001-8455-4622]

JAMES SOWELL [0000-0003-3480-0957]

OSCAR STRANIERO [0000-0002-5514-6125]


\clearpage


\startlongtable
\begin{deluxetable}{ccrrl}
\tabletypesize{\normalsize}
\tablewidth{0pt}
\tablecolumns{6}
\tablecaption{Radial Velocity Observations of IGR~J16194$-$2810}
\label{table:rv_obs}
\tablehead{\colhead{Hel. Julian Date}  & \colhead{Phase} &
\colhead{RV} & \colhead{$(O-C)$\tablenotemark{a}} & 
\colhead{Source\tablenotemark{b}
}
\\
\colhead{HJD$-$2,400,000} & \colhead{} &
\colhead{(km~s$^{-1}$)}  & \colhead{(km~s$^{-1}$)}  & \colhead{} 
}
\startdata
 58230.776 & 0.867 &    14.1 &     1.4 &  IGRINS  \\
 58559.916 & 0.577 & $-$27.0 &     0.0 &  Phoenix \\
 58660.726 & 0.100 &    18.0 &     1.7 &  Fair   \\
 58661.726 & 0.106 &    16.7 &     0.9 &  Fair   \\
 58981.836 & 0.769 &  $-$2.5 &  $-$1.1 &  Fair   \\
 58982.836 & 0.774 &  $-$1.6 &  $-$1.1 &  Fair   \\
 59021.726 & 0.976 &    21.7 &     0.8 &  Fair   \\
 59022.726 & 0.981 &    19.8 &  $-$1.2 &  Fair   \\
 59023.726 & 0.986 &    21.3 &     0.2 &  Fair   \\
 59024.726 & 0.992 &    21.8 &     0.6 &  Fair   \\
 59025.726 & 0.997 &    20.7 &  $-$0.5 &  Fair   \\
 59026.726 & 0.002 &    20.5 &  $-$0.7 &  Fair   \\
 59029.726 & 0.018 &    20.8 &  $-$0.3 &  Fair   \\
 59031.726 & 0.028 &    20.5 &  $-$0.3 &  Fair   \\
 59296.789 & 0.405 & $-$25.6 &  $-$0.1 &  Fair   \\
 59302.984 & 0.437 & $-$26.6 &     1.4 &  Fair   \\
 59303.984 & 0.443 & $-$28.7 &  $-$0.4 &  Fair   \\
 59304.984 & 0.448 & $-$28.6 &     0.0 &  Fair   \\
 59307.984 & 0.463 & $-$28.6 &     0.7 &  Fair   \\
 59308.984 & 0.469 & $-$28.4 &     1.0 &  Fair   \\
 59315.925 & 0.505 & $-$29.8 &     0.1 &  Fair   \\
 59317.925 & 0.515 & $-$30.4 &  $-$0.6 &  Fair   \\
 59318.925 & 0.520 & $-$29.9 &  $-$0.2 &  Fair   \\
 59340.856 & 0.634 & $-$20.5 &     0.9 &  Fair   \\
 59342.856 & 0.645 & $-$20.6 &  $-$0.5 &  Fair   \\
 59345.856 & 0.660 & $-$16.3 &     1.8 &  Fair   \\
 59346.856 & 0.665 & $-$17.1 &     0.2 &  Fair   \\
 59347.856 & 0.671 & $-$17.1 &  $-$0.5 &  Fair   \\
 59349.856 & 0.681 & $-$15.4 &  $-$0.3 &  Fair   \\
 59350.856 & 0.686 & $-$14.0 &     0.4 &  Fair   \\
 59352.856 & 0.697 & $-$13.0 &  $-$0.2 &  Fair   \\
 59353.826 & 0.702 & $-$11.8 &     0.2 &  Fair   \\
 59354.827 & 0.707 & $-$12.4 &  $-$1.2 &  Fair   \\
 59356.827 & 0.717 &  $-$8.7 &     0.9 &  Fair   \\
 59357.807 & 0.722 &  $-$9.5 &  $-$0.7 &  Fair   \\
 59364.797 & 0.759 &  $-$2.5 &     0.5 &  Fair   \\
 59365.797 & 0.764 &  $-$2.6 &  $-$0.4 &  Fair   \\
 59372.767 & 0.800 &     1.6 &  $-$1.9 &  Fair   \\
 59373.767 & 0.805 &     2.3 &  $-$2.0 &  Fair   \\
 59374.767 & 0.810 &     3.8 &  $-$1.3 &  Fair   \\
 59375.767 & 0.816 &     6.3 &     0.4 &  Fair   \\
 59377.756 & 0.826 &     7.0 &  $-$0.4 &  Fair   \\
 59392.716 & 0.904 &    17.0 &     0.3 &  Fair   \\
 59447.672 & 0.189 &     7.1 &     1.9 &  Fair   \\
 59657.983 & 0.282 & $-$11.0 &  $-$1.6 &  Fair   \\
 59658.983 & 0.287 & $-$11.1 &  $-$0.8 &  Fair   \\
 59663.983 & 0.313 & $-$14.6 &  $-$0.4 &  Fair   \\
 59664.984 & 0.318 & $-$15.3 &  $-$0.3 &  Fair   \\
 59675.984 & 0.375 & $-$23.0 &  $-$0.5 &  Fair   \\
 59676.984 & 0.381 & $-$23.1 &     0.0 &  Fair   \\
 59683.915 & 0.417 & $-$26.3 &     0.2 &  Fair   \\
 59684.915 & 0.422 & $-$27.3 &  $-$0.4 &  Fair   \\
 59690.915 & 0.453 & $-$29.7 &  $-$0.9 &  Fair   \\
 59691.916 & 0.458 & $-$29.7 &  $-$0.6 &  Fair   \\
 59693.916 & 0.469 & $-$28.6 &     0.8 &  Fair   \\
 59694.916 & 0.474 & $-$28.9 &     0.7 &  Fair   \\
 59710.836 & 0.556 & $-$28.3 &     0.0 &  Fair   \\
 59711.836 & 0.562 & $-$28.0 &     0.0 &  Fair   \\
 59712.836 & 0.567 & $-$27.6 &     0.1 &  Fair   \\
 59714.836 & 0.577 & $-$27.6 &  $-$0.6 &  Fair   \\
 59717.806 & 0.593 & $-$25.8 &  $-$0.1 &  Fair   \\
 59719.807 & 0.603 & $-$25.1 &  $-$0.3 &  Fair   \\
 59724.797 & 0.629 & $-$23.2 &  $-$1.2 &  Fair   \\
 59730.797 & 0.660 & $-$17.8 &     0.2 &  Fair   \\
 59731.797 & 0.665 & $-$18.2 &  $-$0.9 &  Fair   \\
\enddata
\tablenotetext{a}{O-C = Observed minus calculated radial velocity}
\tablenotetext{b}{IGRINS = IGRINS at GS, Phoenix = Phoenix at GS,
Fair = Tennessee State University 2~m AST and fiber-fed echelle spectrograph at Fairborn Observatory} 
\tablecomments{This table is available in machine-readable form.}
\end{deluxetable}

\clearpage


\begin{deluxetable}{lc}
\tabletypesize{\normalsize}
\tablewidth{0pt}
\tablecaption{IGR~J16194$-$2810 Spectroscopic Orbital Elements and Related Parameters}
\label{table:orbit}
\tablehead{ \colhead{Parameter} & \colhead{Value}  }
\startdata
$P$ (days)               &  192.47 $\pm$ 0.13 \\
$T_0$ (HJD)              &    2459026.33  $\pm$ 0.33 \\
$K$ (km s$^{-1}$)      &  25.58 $\pm$ 0.16 \\
$\gamma$ (km~s$^{-1}$)    & $-$4.36 $\pm$ 0.13 \\
$a$~sin~$i$ (10$^6$ km) & 67.69  $\pm$ 0.42 \\
f(m) ($M_{\sun}$) & 0.3337 $\pm$ 0.0062  \\
S.E.\tablenotemark{a}  (km~s$^{-1}$) & 0.9 \\
\enddata
\tablenotetext{a}{Standard error of an observation of unit weight} 
\end{deluxetable}

\clearpage


\begin{deluxetable}{lcl}
\tablewidth{0pt}
\tablecaption{Parameters of the IGR~J16194$-$2810 Giant} 
\label{table:parameters}
\tablehead{\colhead{Parameter} & \colhead{Value} & \colhead{Source}
}
\startdata
Distance              &  $2101\pm150$ pc                   & Gaia EDR3          \\
Spectral Type         &  M2 III                           & Literature, TiO      \\
Effective Temperature &  $3700\pm100$ K                  & Sp.Ty., photometry   \\
Luminosity            & 573$^{+72}_{-79}$ L$_\odot$    & Distance, K$_0$ \\
Radius                & $58\pm7$ R$_\odot$         & L, T$_{eff}$ \\
Mass                  & $0.91\pm0.07$ M$_\odot$   & Ellipsoidal, NS mass \\ 
Surface gravity\tablenotemark{a} &  5 -- 11 cm s$^{-2}$ & Mass and radius    \\
Inclination           &  55 -- 70$^\circ$                        & Ellipsoidal \\
Equatorial Rotation   &  $18\pm2$ km s$^{-1}$              & $v\,sin\,i$ and inclination  \\
\enddata
\tablenotetext{a}{log(g)$\sim$1}   
\end{deluxetable}

\clearpage


\begin{deluxetable}{lrcl}
\tablewidth{0pt}
\tablecaption{Abundances of the IGR~J16194$-$2810 Giant}
\label{table:abundances}
\tablehead{\colhead{Element} & \colhead{Value} & \colhead{Number of} & \colhead{Species}\\
\colhead{} & \colhead{} & \colhead{lines synthesized} & \colhead{}
}
\startdata
 $[$Fe/H$]$                      & $-$0.14$\pm$0.12  & 7  & Fe I                       \\
 $[$C/Fe$]$                      & $-$0.19$\pm$0.11  & 6  & $^{12}$CO $\Delta$v=2      \\
 $[$N/Fe$]$                      & 0.31$\pm$0.11    & 9 & CN Red $\Delta$v=$-$1      \\
 $[$O/Fe$]$                      & 0.14$\pm$0.13   & 15 & OH $\Delta$v=2            \\
$^{12}$C/$^{13}$C           & 22$\pm$3      & 5 & $^{13}$CO $\Delta$v=2      \\
\enddata
\end{deluxetable}

\clearpage


\begin{deluxetable}{lccc}
\tablewidth{0pt}
\tablecaption{Long Orbital Period Pulsars}
\label{table:pulsar_table}
\tablehead{
\colhead{Name} & \colhead{Orbital Period\tablenotemark{a}} & \colhead{Spin Period}    & \colhead{WD type\tablenotemark{b}} \\
\colhead{}     & \colhead{(days)}         & \colhead{(milliseconds)} & \colhead{}
}
\startdata
J0214$+$5222  & 512.0    & 24.5   & He      \\  
J0407$+$1607  & 669.1    & 25.7   & He      \\  
J0823$+$0159  & 1232.4   & 864.9  & CO      \\  
J1711$-$4322  & 922.5    & 102.6  & He      \\ 
J1803$-$2712  & 406.8    & 334.4  & He      \\ 
J1822$-$0848  & 286.8    & 834.8  &  He:   \\  
J1840$-$0643  & 944.6    & 35.6   & He     \\ 
J2016$+$1948  & 635.0    & 64.9   & He:    \\   
J2204$+$2700  & 815.2    & 84.7   & He:   \\  
\enddata
\tablenotetext{a}{Low mass systems with orbital periods $>$ 200 days from ATNF Pulsar Catalog Version 1.70.  Globular cluster pulsars and systems
with highly uncertain secondary types are omitted.}
\tablenotetext{b}{Colon indicates uncertain designation.}
\end{deluxetable}

\clearpage


\begin{rotatetable}
\begin{deluxetable}{c|rrr|llr|lllr|llll|llr}
\tabletypesize{\footnotesize}
\tablecaption{FuNS Interacting Binary Evolution Models\tablenotemark{a}}
\label{table:os_1}
\tablehead{
\colhead{Model} 
& \multicolumn{3}{c}{Initial} 
& \multicolumn{3}{c}{AIC} 
& \multicolumn{4}{c}{L$_{M_2}$=2.76} 
& \multicolumn{4}{c}{P=192} 
& \multicolumn{3}{c}{Final}
\\[-6pt]
\colhead{Name} 
& \colhead{M$_1$} & \colhead{M$_2$} & \colhead{P} 
& \colhead{L$_{2}$} & \colhead{M$_{2}$} & \colhead{P}
& \colhead{T$_{e\mkern0muf\!f}$} & \colhead{M$_1$} & \colhead{M$_2$} & \colhead{P} 
& \colhead{T$_{e\mkern0muf\!f}$} & \colhead{L$_{2}$} & \colhead{M$_1$} & \colhead{M$_2$} 
& \colhead{M$_1$} & \colhead{M$_2$} & \colhead{P}
}
\startdata
   G1 & 1.30 & 1.20 & 120.00 & 2.74 & 1.08 & 128.10 & 3.57 & 1.25 & 1.08 & 143.00 & 3.56 & 2.82 & 1.26 & 0.88 & 1.30 & 0.46 & 703.50 \\
   G2 & 1.30 & 1.30 &  90.00 & 2.65 & 1.17 &  95.36 & 3.56 & 1.27 & 0.82 & 180.00 & 3.56 & 2.77 & 1.27 & 0.79 & 1.31 & 0.44 & 616.70 \\
   G3 & 1.30 & 1.40 &  80.00 & 2.62 & 1.24 &  84.85 & 3.56 & 1.27 & 0.77 & 190.50 & 3.56 & 2.76 & 1.27 & 0.77 & 1.32 & 0.44 & 595.40 \\
   G4 & 1.30 & 1.40 &  90.00 & 2.68 & 1.24 &  95.40 & 3.56 & 1.26 & 0.89 & 165.80 & 3.56 & 2.79 & 1.27 & 0.82 & 1.31 & 0.45 & 643.70 \\
   G5 & 1.30 & 1.40 & 100.00 & 2.72 & 1.24 & 105.90 & 3.58 & 1.25 & 1.13 & 130.70 & 3.56 & 2.81 & 1.26 & 0.87 & 1.31 & 0.45 & 689.50 \\
   G6 & 1.30 & 1.50 &  80.00 & 2.65 & 1.30 &  85.61 & 3.56 & 1.27 & 0.81 & 181.00 & 3.56 & 2.77 & 1.27 & 0.79 & 1.32 & 0.44 & 615.30 \\
   G7 & 1.30 & 1.50 &  90.00 & 2.71 & 1.30 &  96.20 & 3.57 & 1.26 & 0.97 & 152.20 & 3.56 & 2.80 & 1.27 & 0.84 & 1.31 & 0.45 & 664.80 \\
   G8 & 1.30 & 1.60 &  70.00 & 2.62 & 1.32 &  77.18 & 3.55 & 1.28 & 0.74 & 199.20 & 3.55 & 2.75 & 1.27 & 0.75 & 1.32 & 0.44 & 577.80 \\
   G9 & 1.30 & 1.60 &  80.00 & 2.68 & 1.32 &  87.85 & 3.56 & 1.26 & 0.86 & 172.80 & 3.56 & 2.78 & 1.27 & 0.81 & 1.31 & 0.45 & 631.20 \\
  G10 & 1.30 & 1.70 &  70.00 & 2.64 & 1.27 &  82.84 & 3.55 & 1.27 & 0.77 & 192.20 & 3.55 & 2.76 & 1.27 & 0.77 & 1.32 & 0.44 & 592.40 \\
\hline 
   J1 & 1.25 & 1.30 &  80.00 & 2.60 & 1.04 &  95.43 & 3.55 & 1.28 & 0.70 & 207.20 & 3.55 & 2.74 & 1.27 & 0.73 & 1.32 & 0.44 & 558.70 \\
   J2 & 1.25 & 1.30 &  90.00 & 2.65 & 1.04 & 107.60 & 3.56 & 1.27 & 0.79 & 185.50 & 3.56 & 2.77 & 1.27 & 0.78 & 1.31 & 0.44 & 604.70 \\
   J3 & 1.25 & 1.40 &  80.00 & 2.63 & 1.06 &  99.61 & 3.55 & 1.27 & 0.75 & 194.60 & 3.56 & 2.76 & 1.27 & 0.76 & 1.32 & 0.44 & 586.10 \\
   J4 & 1.25 & 1.40 &  90.00 & 2.68 & 1.06 & 112.20 & 3.56 & 1.26 & 0.86 & 170.70 & 3.56 & 2.78 & 1.27 & 0.81 & 1.31 & 0.45 & 634.00 \\
   J5 & 1.25 & 1.40 & 100.00 & 2.73 & 1.06 & 124.70 & 3.57 & 1.25 & 1.02 & 145.50 & 3.56 & 2.81 & 1.26 & 0.86 & 1.30 & 0.45 & 679.40 \\
   J6 & 1.25 & 1.50 &  80.00 & 2.66 & 1.05 & 106.70 & 3.56 & 1.27 & 0.80 & 184.10 & 3.56 & 2.77 & 1.27 & 0.78 & 1.31 & 0.44 & 608.30 \\
   J7 & 1.25 & 1.60 &  80.00 & 2.68 & 1.01 & 117.80 & 3.56 & 1.26 & 0.84 & 175.90 & 3.56 & 2.78 & 1.26 & 0.80 & 1.31 & 0.45 & 624.60 \\
   J8\tablenotemark{b} & 1.25 & 1.60 & 100.00 & 2.78 & 1.01 & 147.90 & \nodata & \nodata & \nodata & \nodata  & 3.56 & 2.83 & 1.26 & 0.91 & 1.30 & 0.46 & 723.50 \\
\hline
   H1 & 1.20 & 1.20 & 100.00 &  2.67 & 0.90 &129.20 & 3.56 & 1.26 & 0.77 & 191.00 & 3.56 & 2.76 & 1.26 & 0.76 & 1.31 & 0.44 & 591.80 \\
   H2 & 1.20 & 1.20 & 120.00 &  2.75 & 0.89 &156.10 & 3.56 & 1.25 & 0.89 & 175.70 & 3.56 & 2.80 & 1.25 & 0.85 & 1.30 & 0.45 & 667.10 \\
   H3\tablenotemark{b} & 1.20 & 1.20 & 130.00 &  2.79 & 0.89 &170.10 & \nodata & \nodata & \nodata & \nodata  & 3.56 & 2.81 & 1.25 & 0.89 & 1.30 & 0.46 & 702.50 \\
   H4 & 1.20 & 1.40 & 100.00 &  2.73 & 0.92 &144.70 & 3.56 & 1.25 & 0.92 & 164.20 & 3.56 & 2.80 & 1.26 & 0.83 & 1.30 & 0.45 & 656.00 \\
\enddata
\tablenotetext{a}{Units: mass=M$_\odot$, period=days, luminosity=log(L/L$\odot$), effective temperature=log(T$_{eff}$)}
\tablenotetext{b}{In models J8 and H3, the AIC occurs at log L$>$2.76}
\end{deluxetable}
\end{rotatetable}


\begin{deluxetable}{c|llr|llr}
\tabletypesize{\normalsize}
\tablewidth{0pt}
\tablecaption{Predicted q Values\tablenotemark{a}}
\label{table:q_values}
\tablehead{ 
\colhead{Model} & \multicolumn{3}{c}{L=2.76} & \multicolumn{3}{c}{P=192} \\
[-6pt]
\colhead{Name} & 
\colhead{M$_{1}$} & \colhead{M$_{2}$} & \colhead{q} & 
\colhead{M$_{1}$} & \colhead{M$_{2}$} & \colhead{q} 
}
\startdata
G1 & 1.25 & 1.08 & 0.86 & 1.26 & 0.88 & 0.70 \\
G2 & 1.27 & 0.82 & 0.65 & 1.27 & 0.79 & 0.62 \\
G3 & 1.27 & 0.77 & 0.61 & 1.27 & 0.77 & 0.61 \\
G4 & 1.26 & 0.89 & 0.71 & 1.27 & 0.82 & 0.65 \\
G5 & 1.25 & 1.13 & 0.90 & 1.26 & 0.87 & 0.69 \\
G6 & 1.27 & 0.81 & 0.64 & 1.27 & 0.79 & 0.62 \\
G7 & 1.26 & 0.97 & 0.77 & 1.27 & 0.84 & 0.66 \\
G8 & 1.28 & 0.74 & 0.58 & 1.27 & 0.75 & 0.59 \\
G9 & 1.26 & 0.86 & 0.68 & 1.27 & 0.81 & 0.64 \\
G10 & 1.27 & 0.77 & 0.61 & 1.27 & 0.77 & 0.61 \\
\enddata
\tablenotetext{a}{Units: mass=M$_\odot$, period=days, luminosity=log(L/L$\odot$)}
\end{deluxetable}


\end{document}